\documentclass[11pt]{article} 

\usepackage[utf8]{inputenc} 

\usepackage{amsmath,amsfonts,amssymb}
\usepackage{bbm, dsfont}
\PassOptionsToPackage{hyphens}{url}

\newcommand{\qed}{\hfill $\Box$}
\usepackage{geometry} 
\usepackage{graphicx}
\usepackage{epstopdf}
\usepackage{svg}

\usepackage{color, colortbl}
\usepackage{xcolor}
\usepackage{adjustbox}

\usepackage{url} 

\usepackage{array} 
\usepackage{paralist} 
\usepackage{comment}

\newcommand{\field}[1]{\mathbb{#1}}
\newcommand {\R}{\field{R} }
\newcommand {\N}{ {\field{N}} }

\newcommand {\chfnct}{\mathbbm{1} }

\newtheorem{prop}{Proposition}[section]

\newtheorem{rem}[prop]{Remark}

\newtheorem{cor}[prop]{Corollary}

\usepackage{color}
\usepackage{colordvi}
\definecolor{magenta}{rgb}{.5,0,.5}
\definecolor{black}{rgb}{1.0,1.0,1.0}
\definecolor{magenta}{rgb}{.1,0,.3}
\definecolor{gruen}{rgb}{0.2,0.5,.5}
\definecolor{light}{rgb}{ 0.992, 0.961,  0.902}
\definecolor{Tan}{rgb}{ 0.992, 0.9,  0.902}

\newcommand{\komment}[1]{{}}

\title{Opinion models, data, and politics}
\author{Matthias Gs\"anger\footnote{Institute of Political Science and Sociology,
		Julius-Maximilians-University (JMU), Würzburg, Germany}, Volker H\"osel\footnote{School for Computation, Information and Technology, TU M\"unchen (TUM), Germany}, Christoph Mohamad-Klotzbach\footnote{Institute of Political Science and Sociology,
		Julius-Maximilians-University (JMU), Würzburg, Germany}, Johannes M\"uller\footnote{School for Computation, Information and Technology, TU M\"unchen (TUM), Germany and Institute for Computational Biology, Helmholtz Center Munich, Germany}}
\date{} 

\begin{document}
	\maketitle
	\begin{abstract}
We investigate the connection between Potts (Curie-Weiss) models and stochastic opinion models in the view of the Boltzmann distribution and stochastic Glauber dynamics. We particularly find that the q-voter model can be considered as a natural extension of the Zealot model which is adapted by Lagrangian parameters. We also discuss weak and strong effects continuum limits for the models.\\
We then fit four models (Curie-Weiss, strong and weak effects limit for the q-voter model, and the reinforcement model) to election data from United States, United Kingdom, France and Germany. We find that particularly the weak effects models are able to fit the data (Kolmogorov-Smirnov test), where the weak effects reinforcement model performs best (AIC). \\
The resulting estimates are interpreted in the view of political sciences, and also the importance of this kind of model-based approaches to election data for the political sciences is discussed. 
\end{abstract}

\mbox{keywords:} Opinion dynamics; Potts models; Glauber dynamics; q-voter  model; reinforcement model; weak and strong effects continuum limit; data analysis and model comparison; elections; voting behavior; interdisciplinarity. \par\bigskip

\section{Introduction}
The search for common compromises in a discursive social process is at the heart of all democracies. In these debates, citizens seek their standpoint on the basis of information and news, but also based on discussions with family, colleagues, and acquaintances. Opinion dynamics aims to model 
the basic structure of precisely this process~\cite{Galam2008,Mimkes2006,Castellano2009}.\\
Many opinion dynamics models are constructed on communications graphs, where the nodes represent persons, who only interact with neighboring persons~\cite{Ye2019}. Basically, there are two groups of models: Either individuals are equipped with one of a finite number of opinions (typically pro and contra), or their opinions are characterized by a continuous spectrum of possibilities (typically the interval $[0,1]$, see~\cite{Sirbu2016}). Particularly the models with a continuous state space are often used to investigate whether a consensus can be reached in the long run~\cite{Dong2018,Anderson2019,Bernardo2024}. \\
Most papers do not aim to validate their models in analyzing empirical data in a quantitative way~\cite{Sobkowicz2020}. 
Those papers which do address empirical data mainly focus on the dynamics of two opinions within homogeneous groups and do not use an underlying graph structure~\cite{Chinellato2015,CostaFilho1999,Fortunato2007,Chatterjee2013,Kononovicius2017,Hoesel2019,Braha2017,Fernandez-Gracia2014}, as many interesting aspects, like different interaction patterns or phase transitions, already appear in models describing homogeneous 
populations, and do not require interaction graphs. 
Here we find a parallel in mathematical epidemiology, where is it clear that infections spread via contact graphs, but most quantitative models and methods aiming at the description and prediction of the dynamics of a real-world outbreak are based on models that assume homogeneous  mixing~\cite{Andersson2000,Diekmann2012}. Also in the present work, we find that these simple models are sufficiently rich to meet the structure of empirical data. Technically, the main objective of those more empirically oriented  papers is to obtain the invariant distribution of the underlying stochastic process and then to use this result in order to fit model parameters to data. In that, these papers are able discuss possible  mechanisms that generate striking patterns in data~\cite{CostaFilho1999,Fortunato2007,Chatterjee2013,Kononovicius2017,Hoesel2019}, 
aim to reveal changes in communication patterns in the course of time~\cite{Braha2017}, or address spatial communication distances~\cite{Fernandez-Gracia2014}. 
Not only in political processes, but also in other fields as vaccination hesitancy, opinion models contribute to an adequate description of the underlying communication mechanisms, and in that potentially open up ways to handle (in this case) public health problems~\cite{Funk2010,Mueller2022,Sobkowicz2020}. It is, however, noticeable that all these aforementioned approaches from socio-physics and socio-mathematics up to now only have a small or no echo in the social and political sciences. \par\medskip

In the present paper, we aim to achieve three goals: In the first part, we connect two different approaches to opinion models, one driven by stochastic dynamics, and the other one originating in statistical physics; in the second part, we target on a model comparison, to find out which models are able to explain empirical data. The third part will then discuss the usefulness of mathematical models of this type for political sciences.\\
First part of the paper: If we review the literature, we find two main approaches to describe opinion models: One approach formulates mechanisms about how people change their minds when interacting with others in the form of stochastic dynamics. In the simplest case, the voter model~\cite{Liggett1985}, it is assumed that a person copies  faithfully the opinion of some randomly chosen other person. A slightly more refined version of this idea introduces zealots (also called stubborns or activists) who never change their mind~\cite{Mobilia2003,Palombi2014}, which results in the zealot (or noisy voter) model. In the long run, we find an equilibrium in the opinion dynamics, which is termed stationary distribution. This distribution can be used to analyze empirical data. A completely different approach originates in the statistical physics of spin systems. Herein, not the dynamics of the opinions is considered, but it is assumed that only little is known about the state of the population. E.g., surveys could inform us about the abundance of some opinions. To express this partial  knowledge appropriately, a distribution is constructed which maximizes the Shannon entropy under the constraint of the known information. This distribution is called the Boltzmann distribution.\\
We connect both approaches in identifying a dynamical stochastic model (first approach) which generates a stationary distribution that coincides with a Boltzmann distribution (second approach). In this way, we associate a dynamical model with a statistical physics model and {\it vice versa}. This dynamics is called Glauber dynamics. The advantage of this procedure is the construction of a unifying framework for both approaches.  Based on this framework, we construct the q-voter model~\cite{Castellano2009qVoter,Mobilia2015,Nycza2012} as the Glauber dynamics of a family of models which 
has the zealot model as a basis.\\ 
We aim to apply the models to data. Election data usually aggregate information about a large number of people, as constituencies usually comprise thousands to hundreds of thousands of voters. Therefore, a continuum limit is of interest.  Here we note that the zealot model is identical to the Moran model~\cite{Moran1958}, which forms the basis of population genetics~\cite{Hoesel2020}. In population genetics, two different diffusion limits have been established, the weak and the strong effects limit. These two approaches differ in the assumptions about the scaling of the parameters with respect to the population size. We investigate ways to transfer  this idea to general opinion models. \par\medskip
In the second part, we analyze election data from the United States (US), United Kingdom (UK), France (FRA), and Germany (GER) based on  four opinion dynamics models (the Curie-Weiss model, the weak and strong effects continuum limit of the q-voter model, and additionally the weak effects limit of the reinforcement model~\cite{Mueller2022}). We use this analysis to test the models to find out to what extent they are able to describe the data not only qualitatively but also quantitatively. The central finding is that models derived by a weak effects limit perform much better than their strong version. Potentially, these findings will be useful for future empirical studies. \par\medskip 
In the last part, we again change the focus and turn to the interpretation of our results in the light of political science. In particular, political processes and changes in social interactions potentially leave their traces in the election data, and therewith in the estimated parameters. However, this conjecture can only be confirmed based on in-depth political research. These considerations lead us to another important aspect, which is the question of the extent to which socio-mathematical models of this kind could also be a fruitful instrument as an integral part of political science, or whether the methods, objectives and research questions of socio-mathematics and socio-physics on the one hand and political science on the other are too different.

\section{General structure}
In this section, we first introduce the notation before introducing the two different types of models that we will investigate.\medskip 

We consider a population of $N$ individuals numbered $1,...,N$, where each of the individuals supports either opinion A or opinion B. The opinions are coded by $1$ (for A) and $-1$ (for B).
The state space is given by $\Sigma_2 = \{\pm 1\}^N$, such that the $i$'th component $\sigma_i\in\{\pm 1\}$ 
of state $\sigma\in\Sigma_2$ indicates the opinion of individual $i$. We consider Cannings models~\cite{Cannings1974},
that is, all individuals are exchangeable and the population is homogeneous. Particularly, we do not have an interaction graph, 
respectively the interaction graph is the full graph. We introduce the functions
$$ 
n_+(\sigma) = \sum_{i=1}^N \chfnct(\sigma_i=1),
\qquad 
n_-(\sigma)=N-n_+(\sigma),$$
which count the supporters of opinion A (function $n_+(\sigma)$), respectively the supporters of opinion~B (function $n_-(\sigma)$). Our knowledge about the opinion distribution in the population will be expressed by a random measures 
on $\Sigma_2$. 
Due to the assumption of Cannings models, the random measures necessarily are invariant w.r.t.\ permutations of individuals: two states $\sigma^1,\sigma^2\in\Sigma_2$ with $n_+(\sigma^1)=n_+(\sigma^2)$ have the same probability. \\
 That is, any random measure $Q:\Sigma_2\rightarrow[0,1]$ describing the state of the population induces a random measure $P$ on the state space $V_N:=\{0,..,N\}$, which indicates the  number of opinion-A supporters.  Let $\sigma^{(k)}\in\Sigma_2$ be a given state with $n_+(\sigma^{(k)})=k$, then for combinatorial reasons
 \begin{eqnarray}
   P(k) = {N\choose k}\,\, Q(\sigma^{(k)}).  
 \end{eqnarray}
We will find out later that the binomial coefficient, which appears here for symmetry reasons, plays a distinct role in the theory developed below. \\
In the next two sections, we introduce two very different ways used in the literature to construct random measures for the opinion state of the population, that is, on $\Sigma_2$ respectively $V_N$. The first approach is based on stochastic processes, while the second is  statistical in design. We should keep in mind that -- due to the symmetry discussed -- 
every rate and every function used to define the models can be constructed in such a way that it depends on $\sigma\in\Sigma_2$ only via 
$n_\pm(\sigma)$ and the population size $N$.

\subsection{Opinion process}

A stochastic opinion process 
is a $\Sigma_2$-valued Markov process $\hat\sigma_t$, where  single persons reconsider their 
opinion at rate $\nu$. There is a certain probability that this person indeed changes her mind. These probabilities depend on the opinion distribution in the population; as mentioned above, this dependency is established via $n_\pm(\sigma)$ and not via the fine-structure of the state (which individual is an A- and which individual is a B-supporter). 
For mathematical convenience, but without loss of generality, we assume that the rate for $\sigma_ii$ to switch from $-1$ to $1$ is a function of $n_+(\sigma)$ 
and $N$, while that to switch from $1$ to $-1$ depends on $n_-(\sigma)$ and $N$, 
\begin{subequations}
\begin{eqnarray}
\sigma_i=-1\rightarrow\,\,\,1 & \mbox{ at rate} & 
    \nu f^+(n_+(\sigma);N)\\
\sigma_i=\,\,\,1\rightarrow -1 & \mbox{ at rate} & 
    \nu f^-(n_-(\sigma);N).
\end{eqnarray}
\end{subequations} 
To get a feeling for which terms to use for $f^\pm$, we can look at the voter model. Herein, $\nu$ represents the rate at which a person  reconsiders her opinion, and interacts with some randomly chosen person in the population (with so-called selfing, that is, the person might also choose herself). The functions $f^{\pm}(n_\pm(\sigma))=n_\pm(\sigma)/N$ simply specify the probability of interacting with a person of the other opinion. Below we also  consider other examples, where the probability to change  the mind is slightly more involved, but the overall structure of the terms will be similar. 
All other entries in state $\sigma$ are not affected by a flip of the $i$'th person's opinion. 
\par\medskip 

As mentioned above, the $\Sigma_2$-valued process $\hat\sigma_t$ induces a $V_N=\{0,..,N\}$-valued process $X_t$ via $X_t=n_+(\hat\sigma_t)$. The transition rates of $X_t$ are given by 
\begin{subequations}
\begin{eqnarray}
X_t=k\rightarrow k+1 & \mbox{ at rate} & \nu (N-k)\,f^+(k;N)\\
X_t=k\rightarrow k-1 & \mbox{ at rate} & \nu k\,\,\,\,f^-(N-k;N). 
\end{eqnarray}
\end{subequations}
We call a Markov process with transition rates given in this form an opinion process. 
It is straightforward to determine 
the stationary distribution of an opinion process  if we have no absorbing states. As we aim at a specific notation that parallels the usual notation of Potts models, we  
derive the stationary distribution step by step. In what follows we suppress the dependency of $f^\pm$ on $N$. 

\begin{prop}[stationary distribution]\label{invarGen}
Assume $f^\pm(k)>0$ for $k\in V_N=\{0,..,N\}$. Let furthermore 
$F_\pm:V_N\rightarrow\R$ defined by 
\begin{eqnarray}
F_\pm(0)=0,\qquad 
 F_\pm(k) = \frac 1 k \sum_{\ell=0}^{k-1} \ln(f^\pm(\ell))\quad\mbox{for } k\in V_N\setminus\{0\}.
\end{eqnarray}
Denoting the probability of state $k\in V_N$ in the stationary distribution by $p_k$, we have 
\begin{eqnarray}\label{invMeasureOpDyn}
p_k = \,\frac {e^{-\tilde H(k)}}{\tilde Z},\quad 
\tilde H(k)=-\ln({ N \choose k})-k\, F_+(k) - (N-k) F_-(N-k),\quad 
\tilde Z = \sum_{k=0}^N  e^{-\tilde H(k)}.
\end{eqnarray}
\end{prop}
{\bf Proof: } 
The detailed balance equation for the stationary distribution $p_k=P(X=k)$ yields 
\begin{eqnarray}\label{DetBalEqu}
p_k  \,k\, f^-(N-k) = p_{k-1} (N-k+1)\, f^+(k-1).
\end{eqnarray}
Hence, 
\begin{eqnarray*}
 p_k 
&=& p_0\, \prod_{\ell=1}^{k} \frac{(N-\ell+1) f^+(\ell-1)}{\ell\, f^-(N-\ell)} 
= p_0\, \frac{\prod_{\ell=1}^{k}N-\ell+1}{k!}\,\, \prod_{\ell=1}^{k} \frac{f^+(\ell-1)}{f^-(N-\ell)} \\
&=& p_0 { N \choose k} \prod_{\ell=1}^{k} \frac{f^+(\ell-1)}{f^-(N-\ell)}
= \frac{p_0}
{\prod_{\ell=1}^{N}f^-(N-\ell)}\, { N \choose k} 
\bigg(\prod_{\ell=1}^{k} f^+(\ell-1)\bigg)
\bigg(\prod_{\ell={k+1}}^{N} f^-(N-\ell)\bigg)\\
&=& \frac{p_0}
{\prod_{\ell=1}^{N}f^-(\ell)}\, { N \choose k} 
\bigg(\prod_{\ell=1}^{k} f^+(\ell-1)\bigg)
\bigg(\prod_{\ell={1}}^{N-k} f^-(\ell-1)\bigg)\\
&=& C\, { N \choose k}\, 
\exp\bigg(\sum_{\ell=0}^{k-1}\ln(f^+(\ell)) + \sum_{\ell=0}^{N-k-1}\ln(f^{-}(\ell))\bigg),
\end{eqnarray*}
where $C$ is determined by $\sum_{k=0}^N p_k=1$. 
Together with the definition of $F_{\pm}$, $\tilde H(k)$, and $\tilde Z$, this formula proves the proposition. 
\par\qed
\par\medskip

Interestingly, the corresponding stationary distribution on $\Sigma_2$ can be written as
\begin{eqnarray}\label{indiInvMeasureX}
Q(\sigma) 
= Z^{-1}\, \exp\bigg\{n_+(\sigma)\, F_+(n_+(\sigma)) +n_-(\sigma)\, F_-(n_-(\sigma))\bigg \}
= Z^{-1}\, \prod_{i=1}^N \exp\bigg(\,F_{\sigma_i}(n_{\sigma_i}(\sigma))\,\bigg)
\end{eqnarray}
where $F_{\sigma_i}=F_+$ if $\sigma_i=1$, and $F_{\sigma_i}=F_-$ if $\sigma_i=-1$; similarly for $n_{\sigma_i}$. 
Each individual has an independent contribution $\exp(F_{\sigma_i}(.))$ to the 
probability of state $\sigma$, where $F_\pm(.)$ depend on the global statistics of the state via $n_\pm(\sigma)$.  
That is, $F_{\sigma_i}(n_{\sigma_i}(\sigma))$ can be regarded as the environment of individual $i$, which determines the probability of the opinion that individual $i$ has adopted. It is 
furthermore interesting to observe that the population size $N$ does not explicitly appear 
in the expression $\exp(F_{\sigma_i}(n_{\sigma_i}(\sigma)))$.  Only indirectly the number of the opposite-opinion-supporters 
comes in, as $n_+(\sigma)$ and $n_-(\sigma)$ add up to the given population size $N$. \\
For obvious reasons, we call the functions $F_\pm(.)$ the environmental conditions, or simply the  environments, of the opinion process. 
\par\medskip

We have a degree of freedom in eqn.~\eqref{invMeasureOpDyn}. We might add a real constant $A$ in the exponent, 
$p_k = \tilde Z^{-1}\,{ N \choose k}\, e^{-\tilde H(k)+A}$. 
Then, $\tilde Z$ is still defined as the normalizing constant, guaranteeing that $\sum_{k=0}^Np_k=1$. Thus, also in $\tilde Z$ the term $e^{A}$ appears, such that $A$ cancels out and does not affect the value of $p_k$. Below, in the definition of the Boltzmann distribution, we will find a similar invariance. This invariance will be used later to eliminate singularities appearing in Section~\ref{weakEffectsLimit}, where we investigate the large population limit with weak effects.\\ 
For now let us discuss the freedom given by this invariance  more in detail, and particularly explore the implication for the choice of the environments. If we replace $F_\pm(k)$ by 
$\tilde F_\pm(k)$, 
\begin{eqnarray*}
 \tilde F_\pm (k) = F_\pm (k) + U_\pm(k)/k,\qquad\quad k=1,\ldots,N, 
 \end{eqnarray*}
where we choose $U_\pm(0)=0$ in accordance to $F_\pm (0)=0$, and require $U_\pm(k)$ to satisfy
\begin{eqnarray*}
 U_+(k)+U_-(N-k)=A, \qquad\quad k\in V_N,
 \end{eqnarray*}
then the stationary distribution $p_k$ is not affected. Furthermore, by specifying $U_+(k)$ and $A$ we specify the full freedom we do have at this point. We now go backward and determine transition rates $k \tilde f^{\pm}(N-k)$ that produce the new environments. Hereto we 
use  $\ln(\tilde f^{\pm}(k)) = (k+1)\,\tilde F_\pm(k+1)-k\,\tilde F_\pm(k)=\ln(f^{\pm}(k)) + U_\pm(k+1)-U_\pm(k)$ such that 
$$ \tilde f^{\pm}(k) = f^\pm(k) \frac{e^{U_\pm(k+1)}}{e^{U_\pm(k)}}. $$
Since $U_-(N-k)=A-U_+(k)$  we have \\[4pt]
\begin{adjustbox}{max width=\textwidth} 
$ \displaystyle 
 \tilde f^{-}(N-k) = f^-(N-k) \frac{e^{U_-(N-k+1)}}{e^{U_-(N-k)}} 
= f^-(N-k) \frac{e^{A-U_+(k-1)}}{e^{A-U_+(k)}} 
= f^-(N-k) \frac{e^{U_+(k)}}{e^{U_+(k-1)}}.  $ 
\end{adjustbox}\\[4pt] 
The function $U_+$ indeed cancels out in the detailed balance equation~\eqref{DetBalEqu}. 

\begin{cor}\label{freedomKor} Given the stationary distribution for the rates $k\, f^\pm(N-k)$ respectively environments $F_\pm(k) = \frac 1 k \sum_{\ell=0}^{k-1} \ln(f^{\pm}(\ell))$, the set of all opinion processes generating this stationary distribution is characterized by 
\begin{eqnarray}
 \tilde f^{\pm}(k) = f^\pm(k) \frac{e^{U_\pm(k+1)}}{e^{U_\pm(k)}} \qquad\mbox{resp.}\qquad 
\tilde F_\pm(k) = F_\pm(k) + U_\pm(k)/k,
\end{eqnarray} 
where $U_\pm$ are functions satisfying $U_+(0)=U_-(0)=0$ and 
\begin{eqnarray} 
U_+(k)+U_-(N-k)=A \qquad \mbox{for } k\in V_N
\quad\mbox{and some } A\in\R.
\end{eqnarray}
\end{cor}

\subsection{Potts machinery} 

Next, we introduce Potts models, 
where we again consider 
Potts models only on a full graph. Potts models on a full graph are often termed mean-field or Curie-Weiss models. 
In agreement with the literature, 
we return for the moment to the individual-based formulation of the opinion model, that is, we use $\Sigma_2$ as state space.\\ 
We do not know the state of the population. The knowledge we assume to have is the results of some polls. For example, we could observe/measure the fraction of individuals with opinion $+1$. Consequently, we will know the expected number of persons in state $1$, that is, $n_+(.)$. As the Potts models originate in physics, we follow the tradition and call these polls ``observations'', and accordingly, the function $n_+(.)$ an ``observable''.\\
In general, observables are defined as functions $\hat F:\Sigma_2\rightarrow\R$ without further restrictions. Another example of an observable that is often used is the number of pairs with identical opinions minus the number of pairs with different opinions,
$$ \hat F(\sigma)=\sum_{k=1}^{N-1}\sum_{\ell=k+1}^N \sigma_k\sigma_\ell. $$
This observable incorporates information about correlations in the population. \par\medskip 

Let us assume that we have $m$ observables $\hat F_1,\ldots,\hat F_m$. Our knowledge about the state of the population is restricted to the knowledge of $\hat f_\ell:=\hat F_\ell(\sigma)$, $\ell=1,\ldots,m$. We represent our knowledge, and particularly the absence of complete knowledge about the state,  in the form of a random measure $Q$ on $\Sigma_2$. We of course require that $E(\hat F_\ell(\sigma))=\hat f_\ell$, such that $Q$ does express our partial knowledge appropriately. However, there are many random measures that will satisfy this requirement. We express this lack of complete knowledge by the condition that $Q$ maximizes the Shannon entropy $S(Q)=-\sum_{\sigma\in \Sigma_2}  Q(\sigma)\ln(Q(\sigma))$, under the constraints  $E(\hat F_\ell(\sigma))=\hat f_\ell$. The random measure $Q(.)$ 
we construct in this way is called the Boltzmann distribution.
\par\medskip 
 We restrict ourselves to Cannings models, such that the observables $\hat F_\ell$ only depend on $\Sigma_2$ via $n_\pm(\sigma)$, and thus can be defined as maps $\hat F_\ell:V_N\rightarrow\R$. Also the Boltzmann measure can be  defined directly on $V_N$ via $P(n_+(\sigma))={N\choose n_+(\sigma)}Q(\sigma)$. As $Q(\sigma)=Q(\tilde\sigma)$ if $n_+(\sigma)=n_+(\tilde\sigma)$, this formula defines $P$ consistently. However, as the lack of knowledge still concerns the state of the population in $\Sigma_2$, we do not use the original Shannon entropy for $P$, but measure the entropy for $P$ by the entropy for the associated measure $Q$ on $\Sigma_2$, 
 \begin{eqnarray*}
 S_{\Sigma_2}(P) & := &  -\sum_{\sigma\in \Sigma_2}  Q(\sigma)\ln(Q(\sigma)) \\
 &=& -\sum_{k\in V_N}  {N\choose k} \bigg\{ P(k)\, {N\choose k}^{-1}\,\ \ln(\,\,P(k)\, {N\choose k}^{-1}\,\,)\,\bigg\}\\
 &=& -\sum_{k\in V_N}  P(k)\, \left( \ln(P(k))-\ln({N\choose k})\,\right).
 \end{eqnarray*}

\begin{prop}  \label{BolzmannGen}
Let $m\in\N$ denote the number of observables, and $\hat F_\ell:V_N\rightarrow\R$, $\ell\in\{1,..,m\}$  the observables themselves. The Boltzmann distribution is the distribution  $P:V_N\rightarrow\R_+$ which 
maximizes the Shannon entropy $S_{\Sigma_2}(P) =  -\sum_{k\in V_N}  P(k)\, \left( \ln(P(k))-\ln{N\choose k}\,\right)$ under the constraint  
$E(\hat F_\ell(.))=\hat f_\ell\in\R$. If the Boltzmann distribution  $P$ exists, then 
\begin{eqnarray}\label{BoltzmannHamiltonian}
P(k) = \frac {e^{-H(k)}}{Z},\qquad H(k)=-\ln({ N \choose k})-\sum_{\ell=1}^m \lambda_\ell \hat F_\ell(k),\qquad Z=\sum_{k\in V_N} e^{-H(k)}, 
\end{eqnarray}
where $\lambda_\ell$, $\ell\in\{1,..,m\}$, are Langrange multipliers.
\end{prop}
{\bf Proof: }The proof consists of a short computation (see e.g.~\cite{Nicolao2019a,Krapivsky2017}), based on the standard Lagrangian approach for maximization of $S(P)$ under the constraints $E(\hat F_\ell(.))=\sum_{k\in V_N}\hat F_\ell(k)\,P(k)=\hat f_\ell$, $\ell=1,\ldots,m$, and $\sum_{k\in V_N}\,P(k)=1$. Let $P(.)$ denote the set of probabilities $P(k)\in[0,1]$ for $k\in V_N=\{0,\dots,N\}$. We determine all values $P(k)$, by maximizing  
the function 
$$L(P(.),\lambda_1,..,\lambda_{m+1}) = S_{\Sigma_2}(P)+\sum_{\ell=1}^m \lambda_\ell (\,E(\hat F_\ell(.))-\hat f_\ell\,) + \lambda_{m+1}(\,\sum_{k\in V_N} P(k)\,-\,1\,).$$ 
Fix $\tilde k\in V_N$. If we equate the derivative of $L$ with respect to $P(\tilde k)$ to zero, we find 
$$ 0 = \frac{\partial L(P(.))}{\partial P(\tilde k)} = 
-\ln(P(\tilde k)) +\ln({N\choose \tilde k}) -1 + \sum_{\ell=1}^m \lambda_\ell \hat F_\ell(\tilde k)+\lambda_{m+1} $$
and 
$P(\tilde k) = e^{-H(\tilde k)} \,\, e^{\lambda_{m+1}-1}$. 
We obtain $e^{\lambda_{m+1}-1}=1/Z$ by the condition that the probabilities sum up to $1$, that is, $\sum_{k\in V_N} P(k)\,=\,1$.
\par\qed

\begin{rem} 
(a) In accordance with the literature, the Lagrangian multipliers $\lambda_1,\ldots,\lambda_m$ are not specified to actually determine a Boltzmann distribution that indeed satisfies $E(\hat F_\ell(.))=\hat f_\ell\in\R$, but instead the Lagrangian multipliers are from now considered on as parameters of the Boltzmann distribution.\\
(b) We note that additive constants in the Hamiltonian do not affect the stationary distribution, as these constants appear in a multiplicative way in numerator $e^{-H(k)}$ as well as in the denominator $Z$  
of the stationary distribution. 
We make use of this observation below to get rid of singularities appearing in the weak effects limit. 
\end{rem}

The Curie-Weiss model {\it in sensu stricto} is defined by observables that are polynomials of second order, as we will discuss in Section~\ref{CWmodel}. For the time being (Section~\ref{connectSect} and Section~\ref{LPL}) we allow for more general functions as observables, as this is necessary to obtain and utilize a connection between the Curie-Weiss models and the opinion processes, which we discuss next. 

\subsection{Connection between opinion processes and the Curie-Weiss model} \label{connectSect}

An opinion process has an invariant distribution, and a Potts model a  Boltzmann distribution. In order to connect opinion processes and Potts models, we ask which conditions the observables (Potts models) respectively the environments (opinion processes) need to satisfy such that the invariant distribution of an opinion model coincides with the Boltzmann distribution. If we combine Propositions~\ref{invarGen}  and Proposition~\ref{BolzmannGen}, we find the following corollary.
\begin{cor}
The stationary distribution of an opinion model with population size $N$ and environment $F_\pm:V_N\rightarrow\R$ and the Boltzmann distribution for observables $\hat F_\pm:V_N\rightarrow\R$ coincide, if 
\begin{eqnarray}
\hat F_\pm(k)=k\,F_\pm(k) 
\qquad \mbox{ and } \qquad \lambda_\pm=1. 
\end{eqnarray}
\end{cor}
Note that this corollary only states sufficient but not necessary conditions. Corollary~\ref{freedomKor} allows, for example, to construct more observables that generate the same  distribution. Considering the settings of this corollary for general $\lambda_\pm\in\R$, we observe 
$$ \lambda_\pm \hat F_\pm(k) 
= \lambda_\pm k\,F_\pm(k)
= \sum_{\ell=0}^{k-1} \ln(\,\,f^\pm(\ell)^{\lambda_\pm}\,).$$
A Boltzmann distribution with observables derived from environments of an opinion process is, for general $\lambda_\pm\in \R$, the stationary distribution of the  
opinion process with transition rates 
\begin{subequations}
\begin{eqnarray}
X_t\rightarrow X_t+1 & \mbox{ at rate} & \nu \, (N-k)\,f^+(k)^{\lambda_+}\\
X_t\rightarrow X_t-1 & \mbox{ at rate} & \nu \,\,\,\,k\,\,\, 
f^-(N-k)^{\lambda_-}.
\end{eqnarray}
\end{subequations}
We call this family of opinion models the Glauber family for the given observables/en\-vi\-ron\-ments. Please note that the standard Glauber dynamics for the spin up/spin down mean-field Ising model~\cite{Krapivsky2017} utilized the freedom discussed in Corollary~\ref{freedomKor} in choosing a non-trivial function~$U_+(k)$. 
\par\medskip

\section{Large population limits} \label{LPL}
Below we consider applications of opinion processes to data, where often the 
population size is in the magnitude of $N\approx 10^5$. Therefore, a continuity limit is of interest. As above, we might consider the dynamics (opinion process) and work out a diffusion 
limit, or we might focus on the Boltzmann distribution, and consider a continuum limit for that distribution. The interesting 
point is to understand the connection between the two resulting objects. Herein we note that the 
rates $f^\pm$, the corresponding environments $F_\pm$ respectively  the observables $\hat F_\pm$ incorporate parameters which might scale differently
for large $N$. In that, different assumptions about this scaling yield  the weak and the strong effects limits (please do not confuse the weak effects limit, which 
refers to the scaling of the parameters, with a weak limit, which refers to the topology of convergence). \par\medskip 

We assume throughout the current section that $f^\pm$ and $\hat F_\pm$ depend on $k/N$ for $k\in V_N$, that is, on the share of an opinion $x=k/N$ in the population instead on the absolute number of supporters of a given opinion, 
\begin{eqnarray*}
f^\pm(k;N)\quad& \mbox{ is replaced by } &\quad f^\pm(k/N;N) \quad k\in V_N,\\
\hat F_\pm(k;N)\quad& \mbox{ is replaced by }& \quad \hat F_\pm(k/N;N) \quad k\in V_N
\end{eqnarray*}
where 
$$ f^\pm:[0,1]\times N_+\rightarrow\R, \qquad \hat F_\pm ^\pm:[0,1]\times N_+\rightarrow\R,$$
such that $f^\pm(x;N)$, $F_\pm(x;N)$ are well-defined for all $x\in[0,1]$. 
Since the observables are defined via $f^\pm$, we rewrite them separately, see Section~\ref{stronEffectLimitSect} below. \\
For $N$ given, the old and the new scaling are mathematically equivalent. However, if we aim at a limit $N\rightarrow\infty$, the new scaling is a reasonable and very helpful assumption, that the most commonly used models actually fulfill.

\subsection{Strong effects limit}\label{stronEffectLimitSect}
In the strong effects limit, we assume that $f^\pm(x;N)$ approximate functions ${\mathbf f}^\pm(x)\in C^2$ for $N\rightarrow\infty$, 
such that 
\begin{eqnarray} 
  \lim_{N\rightarrow\infty} f^\pm(x;N) = {\mathbf f}^\pm(x) \mbox{ in } C^2[0,1]
  \end{eqnarray} 
is well-defined. For clarity of notation, we write the limiting of rate functions, environments, and observables for $N\rightarrow\infty$ in bold. To adapt the definition region of the associated environments from the discrete state space $V_N$ to the continuous state space $x\in[0,1]$, we re-define the environments as  
$$ F_\pm:[0,1]\times \N\rightarrow\R,\quad F_\pm(x;N)  :=  \frac 1 {N\,x} \sum_{\ell=0}^{\lfloor N\,x \rfloor-1} \ln(f^\pm(\ell/N;N)).$$
For $x=k/N$, $k\in V_N$, we get back the environments as defined in
Proposition~\ref{invarGen}. Therewith, also the environments satisfy a proper limit for $N\rightarrow\infty$, 
\begin{eqnarray} \label{limitEins}
{\mathbf F}_\pm(x):= \lim_{N\rightarrow\infty} F_\pm(x;N)  
= \lim_{N\rightarrow\infty} \frac 1 x \sum_{\ell=0}^{\lfloor N\,x \rfloor-1} \ln( f^\pm(\ell/N;N))\frac 1 N = 
x^{-1}\,\int_0^x \ln({\mathbf f}^\pm(y))\, dy.
\end{eqnarray} 
We also introduce a limit for the observables. Here, a certain subtlety appears: We have $\hat F(k/N;N)=kF(k/N;N)$. As $x=k/N$, 
this formula becomes $\hat F(k/N;N)=N\,xF(x;N)$. In leading order, the observables are ${\cal O}(N)$. We thus  scale  them by $1/N$, and define 
\begin{eqnarray} \label{limitZwei}
\hat {\mathbf F}_\pm(x) 
:= 
\lim_{N \rightarrow \infty} \frac 1 N \hat F(k/N;N) 
=x{\mathbf F}_\pm(x) = \int_0^x \ln({\mathbf f}^\pm(y))\, dy.
\end{eqnarray} 
We emphasize at this point that, when using $\hat {\mathbf F}_\pm(x)$ below, we need to take the scale $1/N$,  introduced at this point, into account.\par\medskip 

To obtain the behavior of the opinion process under the strong effects limit, we briefly sketch 
the Kramers Moyal~\cite{Gardiner2009} expansion of the model, which is -- as usual -- truncated at the second order to obtain a Fokker-Planck (or Kolmogorov forward) equation. 
\begin{prop} The Kramers Moyal expansion up to second order for the Glauber family with 
limiting observables $\hat {\mathbf F}_\pm(x)=\int_0^x \ln({\mathbf f}^\pm(y))\, dy$ is given by 
\begin{eqnarray}
\partial_t u(x,t) &=& - 
\nu\, \partial_x\bigg(\, \bigg((1-x)\, {\mathbf f}^+(x)^{\lambda_+}-x\,{\mathbf f}^-(1-x)^{\lambda_-}\,\bigg)\, u(x,t)\bigg) \label{KramersMoyalStrong}\\
&&+ \frac{\nu}{2N}\, \partial_{xx}\bigg(\, \bigg((1-x)\, {\mathbf f}^+(x)^{\lambda_+}+x\, {\mathbf f}^-(1-x)^{\lambda_-}\,\bigg)\, u(x,t)\bigg).\nonumber
\end{eqnarray}
\end{prop}
{\bf Proof: }
We start with the master equations
\begin{eqnarray*}
 \dot p_k &=& -\nu\,\bigg( (N-k)f^+(k/N;N)^{\lambda_+}+ k\, f^-(1-k/N;N)^{\lambda_-}\bigg) p_k\\
&& + \nu (N-(k-1))f^+(\,(k-1)/N)^{\lambda_+}p_{k-1} + \nu (k+1)f^-(1-(k+1)/N)^{\lambda_-}p_{k+1}.
\end{eqnarray*}
If we now assume that $p_k\approx h\,u(x,t)$ for some smooth probability density $u(x,t)$, where $x=k\, h$ and $h=1/N$, we have $\partial_tu(x,t)\approx\dot p_k$, and 
\begin{eqnarray*}
\partial_t h\,u(x,t) & = &-\nu\, h^{-1} ((1-x)f^+(x;N)^{\lambda_+}+xf^-(1-x;N)^{\lambda_-})\, u(x,t)\\
&& + \nu\, h^{-1}\, (1-x-h)\, f^+(x-h;N)^{\lambda_+}u(x-h,t)\\
&& + \nu\, h^{-1}\, (x+h)\, f^-(1-x-h;N)^{\lambda_-} u(x+h,t).
\end{eqnarray*}
Taylor expansion of the last two terms up to second order, neglecting the error term, and using the limit of $f^\pm(\,.\,;N)$ for $N\rightarrow\infty$ yields
the Kramers Moyal expansion. 
\par\qed
\par\medskip 

Next, we turn to the Boltzmann distribution. In the continuum limit, we will denote the Boltzmann distribution (and later also the stationary distribution of a limiting stochastic opinion process) by $\varphi(x)$. 

\begin{prop}\label{StrongEffectBoltzmann}
The Boltzmann distribution for observables $\hat {\mathbf F}_\pm(x)=\int_0^x \ln({\mathbf f}^\pm(y))\, dy$  and large $N$ is given in leading order by the Hamiltonian
\begin{eqnarray}
 H(x) = N\bigg( - H_2(x) -\lambda_+ \int_0^x \ln({\mathbf f}^+(y))\, dy-\lambda_- \int_0^{1-x} \ln({\mathbf f}^-(y))\, dy\bigg), 
 \label{InvMeasureStrongBoltzmann}
 \end{eqnarray}
where $H_2(x)=-x\ln(x)-(1-x)ln(1-x)$ is the binary entropy. The limiting Boltzmann distribution reads $\varphi(x)=e^{-H(x)}/Z(x)$, where $Z(x)=\int_0^1 e^{-H(x)}\, dx$.
\end{prop}
The result is a consequence of eqn.~\eqref{BoltzmannHamiltonian}, 
the scale of $\hat {\mathbf F}_\pm$ w.r.t.\ $N$ introduced in \eqref{limitZwei}, and the well-known approximation of the binomial coefficient by means of the binary entropy
\begin{eqnarray} \label{BinEntropyApprox}
 \ln({N\choose n}) = 
NH_2(n/N)-\frac 1 2 \ln(2\pi N) - \frac 1 2 \ln(x(1-x))+  {\cal O}(N^{-1}).
\end{eqnarray}

The connection between the stationary distributions of the Kramers Moyal expansion~eqn.~\eqref{KramersMoyalStrong} and the stationary distribution eqn.~\eqref{InvMeasureStrongBoltzmann} is not clear - though they are derived from the associated Potts models and opinion processes, they look rather different. The next proposition clarifies the connection.

\begin{prop} Assume that $N$ is large. Let ${\mathbf f}(x)=(1-x)\,{\mathbf f}^+(x)^{\lambda_+}-x\,{\mathbf f}^-(1-x)^{\lambda_-}$, 
and assume ${\mathbf f}(\mu)=0$, ${\mathbf f}'(\mu)<0$ for some $\mu\in(0,1)$.  Then, the local 
normal approximation of the Boltzmann distribution~\eqref{InvMeasureStrongBoltzmann} 
coincides with the stationary distribution of the Ornstein-Uhlenbeck approximation 
of the Kramers Moyal expansion~\eqref{KramersMoyalStrong} at $x=\mu$ in leading order in $N$. 
\end{prop}
{\bf Proof: } We first investigate the Boltzmann distribution. Since \\[4pt]
\begin{adjustbox}{max width=\textwidth}
$\displaystyle
 \frac{H'(\mu)}{N} = \ln(\mu)-\ln(1-\mu) -\lambda_+\ln({\mathbf f}^+(\mu))+\lambda_-\ln({\mathbf f}^-(1-\mu))
= \ln\bigg(\frac{\mu\,{\mathbf f}^-(1-\mu)^{\lambda_-}}{(1-\mu)\,{\mathbf f}^+(\mu)^{\lambda_+} }\bigg)=0$
\end{adjustbox}\\[4pt]
we have a critical point of H(x) at $x=\mu$. The second derivative reads, again using ${\mathbf f}(\mu)=0$, 
{\small 
\begin{eqnarray*}
&&\frac{H''(\mu)}{N}\\
&=& 
\frac{(1-\mu)\,{\mathbf f}^+(\mu)^{\lambda_+}}{\mu\,{\mathbf f}^-(1-\mu)^{\lambda_-}}\,
\frac{
(1-\mu)\,{\mathbf f}^+(\mu)^{\lambda_+} \frac d{d\mu} (\mu\,{\mathbf f}^-(1-\mu)^{\lambda_-})
-
\mu\,{\mathbf f}^-(1-\mu)^{\lambda_-} \frac d {d\mu} ((1-\mu)\,{\mathbf f}^+(\mu)^{\lambda_+})
}{((1-\mu)\,{\mathbf f}^+(\mu)^{\lambda_+})^2}\\
&=&
\frac{
 \frac d{d\mu} (\mu\,{\mathbf f}^-(1-\mu)^{\lambda_-})
-
\frac d {d\mu} ((1-\mu)\,{\mathbf f}^+(\mu)^{\lambda_+})
}{(1-\mu)\,{\mathbf f}^+(\mu)^{\lambda_+}}
=
\frac{- {\mathbf f}'(\mu)}{(1-\mu)\,{\mathbf f}^+(\mu)^{\lambda_+}}
=
\frac{|{\mathbf f}'(\mu)|}{(1-\mu)\,{\mathbf f}^+(\mu)^{\lambda_+}}.
\end{eqnarray*}}
Hence, $H(x) = H(\mu)+\frac 1 2 (x-\mu)^2 H''(\mu) + {\cal O}((x-\mu)^3)$. Locally, at $x=\mu$, the 
stationary distribution $e^{-H(x)}/Z$ behaves as $N(\mu,\sigma^2)$ with 
$$\sigma^2 = H''(\mu)^{-1}=\frac{(1-\mu)\,{\mathbf f}^+(\mu)^{\lambda_+}}{N\,|{\mathbf f}'(\mu)|}
 = \frac{\mu\,{\mathbf f}^-(\mu)^{\lambda_-}}{N\,|{\mathbf f}'(\mu)|}
.$$
Next, we proceed to the stationary distribution based on the Kramers Moyal expansion~\eqref{KramersMoyalStrong}. The leading order terms of the linearization of 
the drift and the noise term at $x=\mu$ yield the Ornstein-Uhlenbeck approximation 
\begin{eqnarray}
\partial_t u(x,t) &=& - 
\nu\, \partial_x\bigg(\, (x-\mu){\mathbf f}'(\mu)\, u(x,t)\bigg)
+ \frac{\nu}{N}\, \partial_{xx}\bigg(\, (1-\mu)\, {\mathbf f}^+(\mu)^{\lambda_+}\, u(x,t)\bigg)\label{KramersMoyalStrongOU}\\
&=& \partial_x\bigg\{
\nu\, \bigg(\, (x-\mu){\mathbf f}'(\mu)\, u(x,t)\bigg) 
+ \frac{\nu}{N}\, \partial_{x}\bigg(\, (1-\mu)\, {\mathbf f}^+(\mu)^{\lambda_+}\, u(x,t)\bigg)
\bigg\}\nonumber
\end{eqnarray}
where we used  $(1-\mu)\, {\mathbf f}^+(\mu)^{\lambda_+}=\mu\,{\mathbf f}^-(1-\mu)^{\lambda_-}$ in the noise term.
In order to identify the stationary distribution, we substitute $u(x,t)=\varphi(x)$ with $\varphi(x)=e^{-a(x-\mu)^2}$ into the term bracketed with curly brackets,
\begin{eqnarray*}
&&
\nu\, \bigg(\, (x-\mu){\mathbf f}'(\mu)\, e^{a(x-\mu)^2}\bigg) 
+ \frac{\nu}{N}\, \partial_{x}\bigg(\, (1-\mu)\, {\mathbf f}^+(\mu)^{\lambda_+}\,e^{-a(x-\mu)^2}\bigg)\\
&=&
\bigg\{
{\mathbf f}'(\mu)\,  
-2a \frac{\nu}{N}\, (1-\mu)\, {\mathbf f}^+(\mu)^{\lambda_+}\,
\bigg\} (x-\mu)\,e^{-a(x-\mu)^2}.
\end{eqnarray*}
This term becomes zero, and therewith $\varphi(x)$ an invariant measure, if $a^{-1}=2\frac{(1-\mu)\,{\mathbf f}^+(\mu)^{\lambda_+}}{N\,|{\mathbf f}'(\mu)|}$. Therefore, 
the stationary distribution of this
approximate Fokker-Planck equation is a normal distribution $N(\mu,\sigma^2)$, where 
$$\sigma^2 = \frac{(1-\mu)\,{\mathbf f}^+(\mu)^{\lambda_+}}{N\,|{\mathbf f}'(\mu)|}
 = \frac{\mu\,{\mathbf f}^-(1-\mu)^{\lambda_-}}{N\,|{\mathbf f}'(\mu)|}
.$$
 \par\qed
\par\medskip

For $N$ large, the stationary distribution will be concentrated close to $\mu$, and hence in the relevant regions 
both stationary distributions, that of eqn.~\eqref{KramersMoyalStrongOU} and of eqn.~\eqref{InvMeasureStrongBoltzmann}, coincide. In that, we consider the Kramers Moyal expansion~\eqref{KramersMoyalStrong} as the Glauber dynamics of the Boltzmann distribution.

\subsection{Weak effects limit}\label{weakEffectsLimit}
The idea of the weak effects limit is to take a simple reference model as a basis and to perturb this model in such a way that for $N\rightarrow\infty$ the transition rates converge back to that of the reference model. For good reasons, as we will find out later, we use the voter model as the reference model. In the voter model, at rate $\nu$ a person copies the opinion of a randomly chosen person in the population (inclusive ``selfing'', that does mean that the focal person might by chance copy her  opinion from herself). Therewith, 
\begin{subequations}
\begin{eqnarray}
X_t=k\rightarrow k+1 & \mbox{ at rate} & \nu (N-k)\,f_{voter}^+(k/N)=\nu (N-k)\,\frac k N\\
X_t=k\rightarrow k-1 & \mbox{ at rate} & \nu k\,\,\,\, f_{voter}^-(1-k/N)=\nu k\,\,\,\,\frac{N-k} N, 
\end{eqnarray}
\end{subequations}
and hence, $f_{voter}^\pm(x)=x$. For the weak effects limit, we allow for rates $f^\pm(x;N)$ which depend on $N$, as long as $\lim_{N\rightarrow\infty}f^\pm(x;N)=f_{voter}^\pm(x)$. Also the Lagrangian parameters are allowed to depend on $N$, $\lambda_\pm=\lambda_\pm(N)$, and the paradigm of weak effects requires again 
$\lim_{N\rightarrow\infty}f^\pm(x;N)^{\lambda_\pm(N)}=f_{voter}^\pm(x)$. Therefore, the expansion of $f^\pm$ w.r.t.\ $1/N$ has $f_{voter}^\pm(x)$ as zero order term, and some arbitrary (well-behaved) function $g^\pm(x)$ as first order coefficient. Similarly, the zero order term of the expansion of $\lambda^\pm(N)$ is $1$, while the first order terms are some parameters $\kappa^\pm\in\R$, which we are free to choose. All in all, the Glauber family suited for the weak effects limit assumes the form
\begin{eqnarray}\label{weakEffectRates}
f^\pm(k/N;N)^{\lambda_\pm(N)}
= 
\bigg(f_{voter}^\pm(k/N) + \frac 1 N g^\pm(k/N)+{\cal O}(N^{-2})\bigg)^{1+\kappa_\pm/N},     
\end{eqnarray}
As we will see, for the weak effects limit, we not only assume the appropriate scaling of the parameters, but also rescale time. In that, even for $N\rightarrow\infty$, we still obtain a non-trivial limiting process and do not simply return to the voter model.\\
As a consequence, we need to re-consider the large population limit done in \eqref{limitEins} and \eqref{limitZwei} more in detail, and also pay attention to terms of order ${\cal O}(N^{-1})$. 
\begin{prop}
Consider the observables for the opinion process defined by \eqref{weakEffectRates}, 
$$\hat F_\pm(x;N) = \sum_{\ell=0}^{\lfloor N\,x \rfloor-1} 
\ln\bigg(\frac \ell N + \frac 1 N g^\pm(\ell/N)+{\cal O}(N^{-2})\bigg)\,\frac 1 N.$$
With the definition 
\begin{eqnarray}
     G^\pm(x) =  \int^x g^\pm(y) /y \, dy.
\end{eqnarray}
the leading order terms of the expansion of $\hat F_\pm(x;N)$ in $1/N$ reads 
$$ \hat F_\pm(x;N) = x\,(\ln(x)-1)  -\frac 1 {2\,N} \ln(x)  + \frac 1 N  G^\pm(x)+{\cal O}(N^{-2}). $$
\end{prop}
{\bf Proof:}  We first rewrite the sum such that it extends to $\lfloor N\,x \rfloor$ instead of $\lfloor N\,x \rfloor-1$,
\begin{eqnarray*}
 \hat F_\pm(x;N) &=&    \sum_{\ell=0}^{\lfloor N\,x \rfloor-1} 
\ln\bigg(\frac \ell N + \frac 1 N g^\pm(\ell/N)+{\cal O}(N^{-2})\bigg)\,\frac 1 N\\
&=& \sum_{\ell=0}^{\lfloor N\,x \rfloor} 
\ln\bigg(\frac \ell N + \frac 1 N g^\pm(\ell/N)+{\cal O}(N^{-2})\bigg)\,\frac 1 N\\
&&  
\qquad-\ln\bigg(\frac {\lfloor N\,x \rfloor} N+ \frac 1 N g^\pm(\lfloor N\,x \rfloor/N)+{\cal O}(N^{-2})\bigg)\,\frac 1 N.
\end{eqnarray*}
Next, we replace the sum by an integral. 
Here we take the Euler-McLaurin correction terms into account in the step from sum  to integral. Furthermore, we note that an additive constant in the Hamiltonian does not change the stationary distribution. 
Instead of $\sum_{\ell=0}^{\lfloor N x\rfloor}(\ldots)$ we can change the lower starting value to any value independent on $x$, and might e.g.\ consider $\sum_{\ell=\lfloor N/2\rfloor}^{\lfloor N x\rfloor}(\ldots)$ instead. Only the upper bound of the sum matters: We only need an anti-derivative. To express this fact, we skip the lower bound of the integral and proceed (where the first equal sign has to be interpreted with the knowledge that we did drop some irrelevant term)
\begin{eqnarray*}
 \hat F_\pm(x;N) &=&   \int^{\lfloor N\,x \rfloor} 
\ln\bigg(\frac \ell N + \frac 1 N g^\pm(\ell/N)+{\cal O}(N^{-2})\bigg)\, d\ell\,\frac 1 N \\
&&
\qquad  -  \frac 1 2 
\ln\bigg(\frac {\lfloor N\,x \rfloor} N + \frac 1 N g^\pm((\lfloor N\,x \rfloor)/N)+{\cal O}(N^{-2})\bigg)\,\frac 1 N +{\cal O}(N^{-2})\\
%
 &=&  \int^x \ln\bigg(y + \frac 1 N g^\pm(y)\bigg)\, dy   
 - \frac 1 2 
\ln\bigg(x + \frac 1 N g^\pm(x)\bigg)\,\frac 1 N +{\cal O}(N^{-2})\\
&=&    \int^x \ln\bigg(y + \frac 1 N g^\pm(y)\bigg)\, dy  -  \frac {\ln(x)} {2 N} - 
\frac 1 2  \ln\bigg(1 + \frac 1 N g^\pm(x)/x\bigg)\,\frac 1 N 
+{\cal O}(N^{-2})\\
&=& \frac {-1} {2\,N} \ln(x)  + \int^x \ln\bigg(y\,\,\bigg\{1 + \frac 1 {Ny} g^\pm(y)\bigg\}\bigg)\, dy    +{\cal O}(N^{-2})\\
&=& x(\ln(x)-1)-\frac 1 {2\,N} \ln(x)  +   \int^x \ln\bigg(1 + \frac 1 {Ny} g^\pm(y)\bigg)\, dy    +{\cal O}(N^{-2}).
\end{eqnarray*}
Note that 
$\lim_{N\rightarrow\infty} \int^x \ln\bigg(1 +\frac 1 {y\,N} g^\pm(y)\bigg)\, dy=0$ (in the sense that $0$ is a possible limiting anti-derivative), such that this integral only contributes to terms of order ${\cal O}(N^{-1})$ or higher. We introduce 
$$ G^\pm(x) = \lim_{N\rightarrow\infty} N \int^x \ln\bigg(1 +\frac 1 {y\,N}  g^\pm(y)\bigg)\, dy 
= \int^x g^\pm(y) /y \, dy
$$
 \par\qed
\par\medskip

Therewith, we are in the position to establish the following proposition.

\begin{prop}
Assume that the functions $f^\pm(x)$ scale with $N$  as described above,
$$f^\pm(x;N)
= 
\bigg(x + \frac 1 N g^\pm(x)+{\cal O}(N^{-2})\bigg)$$
and scale the Lagrangians by $\lambda_\pm = 1+\kappa_\pm/N$. 
Then, in leading order, the Hamiltonian reads
\begin{eqnarray}
H(x) &=& 
\ln(x(1-x))  - G^+(x) - G^-(1-x)-\kappa_+\,\zeta(x)-\kappa_-\,\zeta(1-x) \label{weakHamiltonian}
\end{eqnarray}
with $\zeta(x)=-x(1-\ln(x))$.
\end{prop}
{\bf Proof: } Recall $H_2(x)=-x\ln(x)-(1-x)\ln(1-x)$. With the scaling  $\lambda_\pm=1+\kappa_\pm/N$, we obtain 
\begin{eqnarray*}
&&\lambda_+\hat F_+(x; N)+\lambda_-\hat F_-(x;N)\\
&=& - N\, H_2(x)  - N  
- 
\frac{1}{2} \ln(x(1-x))  +\kappa_+x\ln(x)+\kappa_-(1-x)\ln(1-x)\\
&&-\kappa_+ x-\kappa_-(1-x) +G^+(x)+G^-(1-x)+{\cal O}(N^{-1}).
\end{eqnarray*}
We use the approximation of the binomial coefficient~\eqref{BinEntropyApprox} to obtain the result with 
$H(x) = -\ln{N\choose {N x}} - \lambda_+\hat F_+(x;N)-\lambda_-\hat F_-(x;N)$, where we drop terms 
independent of $x$ and terms of higher order in $N^{-1}$. 
\par\qed
\par\medskip

It is remarkable and typical for the weak effects scaling that the Hamiltonian, and in that also the Boltzmann distribution, 
becomes independent of~$N$. 
We now turn to the underlying opinion process, and discuss the Kramers Moyal expansion under the scaling assumed. 

\begin{prop}
The Kramers Moyal expansion of the Glauber family under the weak effects-scaling in rescaled time $T=\nu t/N$ is given by {\small 
\begin{eqnarray}\label{weakEffectsFP}
u_T(x,T) &=& -\partial_x\bigg( ((1-x)g^+(x)-x g^-(x)+\kappa_+(1-x)\ln(x)+\kappa_- x\ln(1-x))\, u(x,T)\bigg)\\
&&+ \partial_{xx}\bigg( x(1-x)\, u(x,T)\bigg).\nonumber 
\end{eqnarray}}
\end{prop}
{\bf Proof: } The Glauber family with the weak effects scaling is defined by 
$$ 
f^\pm(k/N)
= 
\bigg(\frac k N + \frac 1 N g^+(k/N)+{\cal O}(N^{-2})\bigg)^{1+\kappa_\pm/N}, 
$$
while the rate $X_t=k\rightarrow  k+1$ reads $\nu (1-x)f^+(..)|_{k=x\,N}$ and that for  $X_t=k\rightarrow k-1$ is  $\nu \,x\,f^-(..)|_{k=(1-x)\,N}$. 
The drift term becomes in leading order {\small 
\begin{eqnarray*}
&&\nu(1-x)\, \bigg(x + \frac 1 N  g^+(x)+{\cal O}(N^{-2})\bigg)^{1+\kappa_+/N}- \nu x 
\bigg((1-x) + \frac 1 N g^-(1-x)+{\cal O}(N^{-2})\bigg)^{1+\kappa_-/N}\\
&=& \frac \nu N \bigg( (1-x)g^+(x)-x g^-(x)+\kappa_+(1-x)x\ln(x)+\kappa_- x(1-x)\ln(1-x) \bigg) + \mbox{h.o.t.}
\end{eqnarray*}}
(where h.o.t.\ is a placeholder for higher order terms) and the coefficient of the noise term becomes in leading order {\small 
\begin{eqnarray*}
&&\nu(1-x)\, \bigg(x + \frac 1 N  g^+(x)+{\cal O}(N^{-2})\bigg)^{1+\kappa_+/N}+ \nu x 
\bigg((1-x) + \frac 1 N  g^+(1-x)+{\cal O}(N^{-2})\bigg)^{1+\kappa_-/N}\\
&=& 2\,\nu\, x(1-x) + \mbox{h.o.t.}
\end{eqnarray*}}
If we rescale time $T=\nu\,t/N$ we obtain the result.
\par\qed
\par\medskip

The Karmers-Moyal expansion in rescaled time becomes, as the Hamiltonian, independent of $N$.
\begin{prop} The stationary distribution of the Kramers Moyal expansion is identical with the Boltzmann distribution $\varphi(x)=\exp(-H(x))/Z$, where the Hamiltonian $H(x)$ is given in~\eqref{weakHamiltonian}.
\end{prop}
{\bf Proof: } We plug the Boltzmann distribution into the right-hand side of the Kramers Moyal expansion. Thereto we note that \\[4pt]
\begin{adjustbox}{max width=\textwidth}
$\displaystyle \varphi'(x) = - \varphi(x)\, H'(x) 
= - \bigg(\frac 1 x - \frac 1 {1-x} - (G^+)'(x)+(G^-)'(1-x) 
- \kappa_+\ln(x) -\kappa_-\ln(1-x)\bigg) \varphi(x).
$
\end{adjustbox}\\[4pt]
Furthermore, $\frac d {dx} G^\pm(x) = g^\pm(x)/x$, such that 
\begin{eqnarray*}
&& \partial_x\bigg(x(1-x)\varphi(x)\bigg) = -x\varphi(x)+(1-x)\varphi(x) + x(1-x)\varphi'(x)\\
&=& \bigg( (1-x) g^+(x)-x g^-(x) 
+ x(1-x)\kappa_+\ln(x) + \kappa_-x(1-x)\ln(1-x)\bigg)\varphi(x).
\end{eqnarray*}
If we take the derivative of this equation w.r.t.\ $x$ we indeed find  that $\varphi(x)$ is a stationary solution of eqn.~\eqref{weakEffectsFP}.
\par\qed\par\medskip

\begin{rem}\label{noGoRem} For the weak effects limit, we have chosen the voter model with $f^\pm(x)=x$ as the reference model. If $N\rightarrow\infty$, the rates of the model at hand converge back to this model. This choice looks, at first glance, arbitrary. It is, however, up to the freedom characterized in Corollary~\ref{freedomKor}, a unique choice: The binary entropy $H_2(x)$ generates in the Hamiltonian terms $Nx\ln(x)$ and $N(1-x)\ln(1-x)$. For a weak effects limit to exist, these terms of order ${\cal O}(N)$ need to be balanced and annihilated by the environments of the reference model, which already forces the reference model to be the voter model (or some model which is, according to Corollary~\ref{freedomKor}, equivalent to the voter model).
\end{rem}

\section{Four models: Curie-Weiss, weak and strong q-voter model, and reinforcement}
\label{CZR}
We use the framework introduced above to briefly introduce four  opinion models we intend to apply to data.

\subsection{Curie-Weiss model}\label{CWmodel}
To introduce the classical Curie-Weiss model, we start with the Potts machinery. Recall that the central ingredients are observables, that is functions $\hat F_\pm: V_N\rightarrow \R$, which form constraints when determining the random measure maximizing the Shannon entropy.  Perhaps the most simple, non-trivial case is given by observables which are polynomials of second order, 
$$  F_\pm(k/N;N)= a_\pm k+b_\pm k^2/N
= N\bigg(a_\pm\,x\,+b_\pm\,x^2\,\bigg)_{x=k/N}.$$
We do not need a zero order term, as additive constants in the Hamiltonian do not influence the Boltzmann distribution. Furthermore, we scale the quadratic term by $1/N$ to balance the squared terms in case of large $N$.  With this setting, the Hamilton defined in \eqref{BoltzmannHamiltonian} reads  
\begin{eqnarray*}
H(x) &=& - \ln{N\choose N x} - \lambda_+ F_+(x;N)  - \lambda_- F_-(1-x;N) \\
&=& -\ln{N\choose N x} - N\,\bigg( \lambda_+b_+\,x^2+\lambda_-b_-\,(1-x)^2 + \lambda_+a_+\,x + \lambda_-a_-\,(1-x)
\bigg).
\end{eqnarray*} 
We might rewrite the quadratic terms as \\[4pt]
\begin{adjustbox}{max width = \textwidth}
$\displaystyle
\lambda_+b_+x^2+\lambda_-b_-(1-x)^2 
= \frac{(\lambda_+b_++\lambda_-b_-)}2 (x^2+(1-x)^2) 
+ (\lambda_+b_+-\lambda_-b_-) x
+\frac{\lambda_-b_- - \lambda_+b_+} 2
$
\end{adjustbox}\\[4pt]
We define $J=\lambda_+b_++\lambda_-b_-$ and $h _\pm$ appropriately (e.g.\ $h_+=\lambda_+b_+-\lambda_-b_-$). 
Furthermore, for historical reasons, we introduce $h_-=0$ and drop terms independent of $x$. Therewith, we obtain 
\begin{eqnarray}
 H(x) = -\ln{N\choose x N}  -N\bigg\{\frac J{2}\, \bigg(x^2+(1-x)^2\bigg) + h_+ x+h_-(1-x)\bigg\},
\end{eqnarray} 
which is the standard form of the model on the state space $V_N$~\cite{Krapivsky2017}. We can still reduce the number of parameters from four ($N,J,h_+,h_-$) to three $(N,J,h)$ with $h=h_+-h_-$ as the additive constant, which appears, can again be dropped.\par\medskip 

Last, we check existence of the strong and the weak effects limit. The strong effects limit can be derived trivially, simply by replacing the binomial coefficient by $N H_2(x)$, cf.~\eqref{BinEntropyApprox},  
$$ H(x) = - N \bigg( H_2(x) +\frac J 2 (x^2+(1-x)^2)+h_+x+h_-(1-x)\bigg).$$
The binary entropy introduces logarithmic terms of order $N$ into the Hamiltonian $H(x)$. As the observables of the Curie-Weiss model consist of polynomial terms, they cannot cancel these logarithmic terms, such that the Hamiltonian always incorporates nontrivial terms of order ${\cal O}(N)$. In the proper weak effects limit, however, terms of this order are not present. Thus, a weak effects limit for the Curie-Weiss model is not possible (cf.~Remark~\ref{noGoRem}). This will be different for the other models we shall discuss next.

\subsection{Two flavors of the q-voter model}

For the q-voter model, we do not start with the stationary distribution but the transition rates. 
Perhaps the most simple extension of the voter model where no opinion can die out is the zealot model, where we have $N^\pm$ zealots for the 
opinion $\pm 1$. Palombi and Toti~\cite[p. 337]{Palombi2014} call zealots "stubborn agents [...] who never change political preference". In our case, zealots are not real persons but represent sources of information that stand for a specific opinion. These could be politicians, friends, newspapers, or social media channels. As in the voter model, individuals copy their opinion from a randomly chosen person, now also from the zealots, which leads to 
$$ f^\pm(k;N) = \frac{N^\pm + k}{N+N^-+N^+}.$$
The corresponding Glauber family is given by 
\begin{subequations}
\begin{eqnarray}
X_t=k\rightarrow k+1 & \mbox{ at rate} & \nu (N-k)\,\left (\,\frac{N^+ + k}{N+N^-+N^+}\right )^{\lambda_+}\\
X_t=k\rightarrow k-1 & \mbox{ at rate} & \nu k\, \left (\,\frac{N^- + (N-k)}{N+N^-+N^+}\right )^{\lambda_-}. 
\end{eqnarray}
\end{subequations}
For $\lambda_+=\lambda_->1$, this is the q-voter model for a homogeneous population~\cite{Castellano2009qVoter}. In the case of $\lambda_\pm=1$, we are back in the zealot model: A person simply copies the opinion of another person or a zealot. If $\lambda_\pm>1$, the model can be interpreted as follows: The person will ask $\lambda_\pm$ other persons for their opinion, and will only change her opinion if all these other persons have the identical opinion. 

\subsubsection{q-voter model -- strong effects}
We consider the strong effects limit: If $N^\pm = \eta^\pm\, N$, that is, if the number of zealots scales linearly with the population size where $\eta^\pm$ are the proportionality constants, 
$$ f^\pm(x;N) = 
\frac{N\,\eta^\pm\,+ N\,x}{N+N\,\eta^-+N\,\eta^+}
=\frac{\eta^\pm + x}{1+\eta^-+\eta^+},$$
such that $ f^\pm(x;N)$ becomes independent of $N$, and $f^\pm(x;N)\equiv {\mathbf f}^\pm(x)$. Therewith, the limiting 
observables are given by \\[4pt]
\begin{adjustbox}{max width=\textwidth}
$\displaystyle \hat {\mathbf F}_\pm(x) =  \int_0^x\ln\left(\frac{\eta^\pm + y}{1+\eta^-+\eta^+}\right)\, dy
= (x+\eta^\pm)\,\ln(\eta^\pm+x)-x
- x\ln(1+\eta^-+\eta^+) +C
$
\end{adjustbox}\\[4pt]
where $C$ is a constant.  
The stationary distribution in the strong effects limit is given by $\varphi(x)=Z^{-1}\exp(-H(x))$, 
where
\begin{eqnarray}
 H(x) &=& -N\,\bigg(H_2(x) 
+\lambda_+\bigg[(x+\eta^+)\,\ln(\eta^++x)-x(1+\ln(1+\eta^-+\eta^+))\bigg]\\
&&\qquad\quad
+\lambda_-\bigg[((1-x)+\eta^-)\,\ln(\eta^-+(1-x))-(1-x)(1+\ln(1+\eta^-+\eta^+))\bigg]
\bigg).\nonumber 
\end{eqnarray}
In the last formula, we again made use of the fact that we are allowed to drop constant terms from the Hamiltonian.

\subsubsection{q-voter model -- weak effects}

We now go into the weak effects limit for the q-voter model. The basis is the zealot model with $ f^\pm(k;N) = \frac{N^\pm + k}{N+N^-+N^-}.$ For the weak effects limit, the number of zealots $N^\pm$ does not scale with the population size, and in that, zealots become rare if $N$ becomes large. 
We can choose the time units, and in this, we have a degree of freedom in the form of a multiplicative positive constant. This freedom can be used to replace the original denominator $N+N^-+N^-$ 
by $N$, and work with 
$$ f^\pm(x;N) = x+\frac 1 N \,g^\pm(x),\qquad  g^\pm(x) \equiv N^\pm. $$
That is, for our particular choice the functions $g^\pm(x)$ are independent of $x$. Recall that we also expand the Lagrangians in terms of $N$ and write $\lambda_\pm=1+\kappa_\pm/N$. 

To obtain the weak effects limit, we  note that $G^\pm(x) = \int^x g^\pm(y)/y\, dy = \ln(x)\, N^\pm$
and obtain the Hamiltonian \\[4pt]
\begin{adjustbox}{max width=\textwidth}
$\displaystyle 
H(x) = -\ln(x)(N^+-1) - \ln(1-x) (N^--1) -\kappa_+x\ln(x)-\kappa_-(1-x)\ln(1-x)
+(\kappa_+-\kappa_-) \, x .
$  
\end{adjustbox}\\[4pt]
with the corresponding  stationary distribution
\begin{eqnarray}
 \varphi(x) = 
 C \,x^{N^+-1}\,(1-x)^{N^--1}\, e^{-\kappa_+x(1-\ln(x))}\,\,e^{-\kappa_-(1-x)(1-\ln(1-x)) }.
 \end{eqnarray}
The invariant distribution becomes a beta distribution in the case of the zealot model ($\kappa_\pm=0$); this result was first 
derived in the context of population genetics, where the zealot model is termed Moran model~\cite[page 108]{Ewens2004}. 
The extension to the weak effects limit of the Glauber family presented here is novel.

\subsection{Reinforcement model -- weak effects}
\begin{figure}
\begin{center}
\includegraphics[width=14cm]{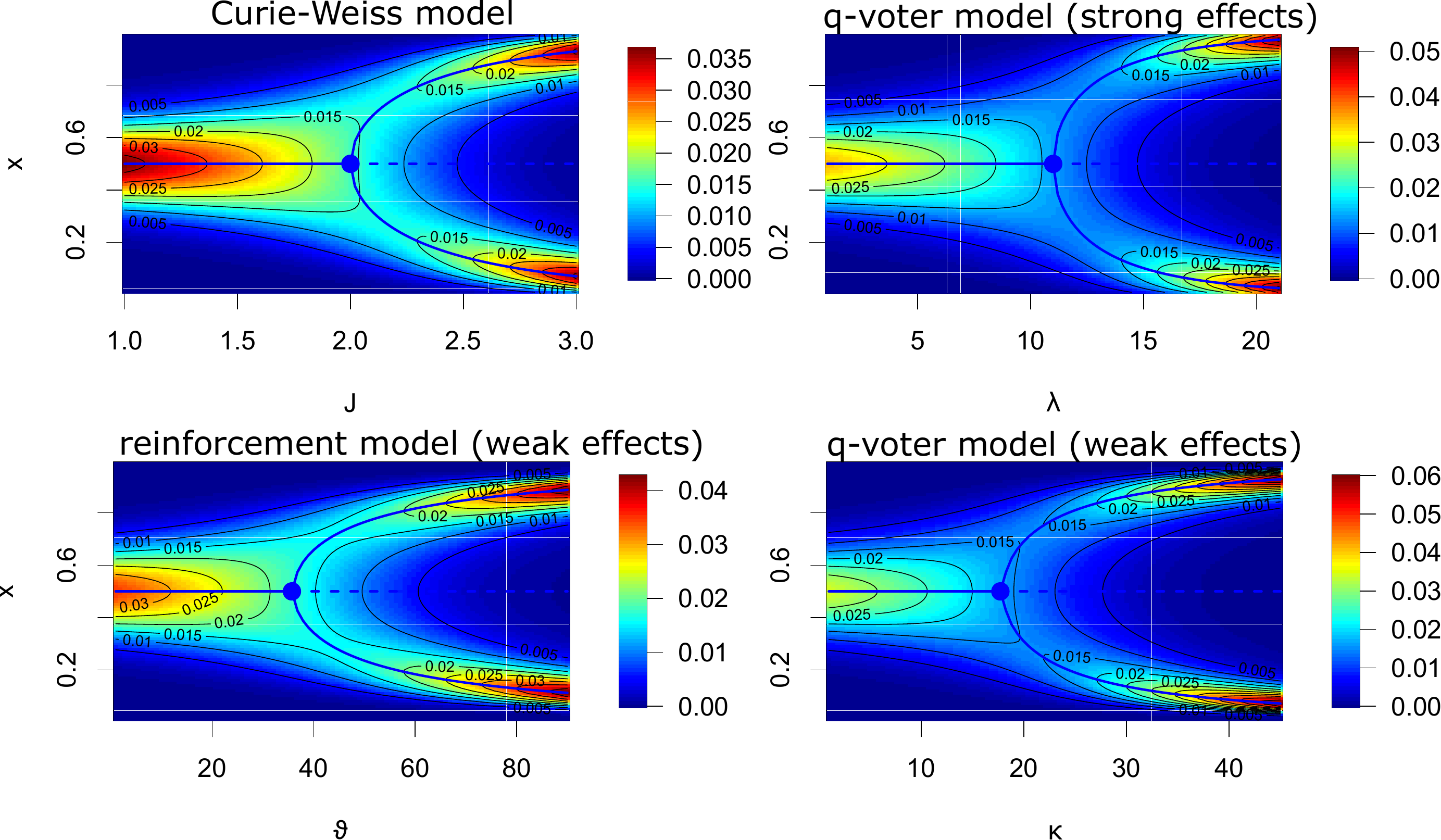}
\end{center}
\caption{Phase transitions of the four models in the symmetric case. For a given parameter (x-axis), the density of the distribution is indicated (by the heat- and contour plot) over $x$ (y-axis). The blue lines indicate local maxima (solid lines) and local minima (dashed line) if the parameter (given at the x-axis) is fixed, while the dot marks the phase transition.
(Curie-Weiss: $h_\pm=0$; strong q-voter:  $\eta_\pm=5$ and $N=20$; weak q-voter:  $N^\pm=10$, $\kappa_\pm=\kappa$; weak reinforcement: $N^\pm=10$,  $\vartheta_\pm=\vartheta$).}\label{bifuFig}
\end{figure}

We add one more model, which also allows for a strong as well as weak effects limit, and which is, as the q-voter model, also a descendant from the zealot model: The reinforcement model~\cite{Mueller2022}. The idea of the reinforcement model is to express the psychological mechanisms which lead to filter bubbles and echo chambers: Several kinds of cognitive biases let  individuals communicate with persons of the opposite opinion with less awareness than with individuals of their own opinion. Some interactions with the opposite group are ignored. In that, the effective size of the opposite group is reduced by a factor $\theta_\pm\in(0,1]$. The zealot model is described by 
$$  f^\pm(k/N;N) = \frac{\theta_\pm(N^\pm + k)}{N^\pm + N-k + \theta_\pm(N^\mp+k)}.$$
We focus on the weak effects limit and hence keep (as in the weak limit of the q-voter model) $N^\pm$ independent of $N$. Furthermore, we choose $\theta_\pm = 1-\vartheta_\pm/N$ such that we return to the voter model if $N$ becomes large; the Lagrangians are taken to be $\lambda_\pm=1$ and are not scaled. It turns out that the computations assume a simpler form (additive constants will vanish below in the first order term of the expansion) if we use a trivial time scale, such that a multiplicative term $N^+/N+1+N^-/N$ appears in the rates, 
$$  f^\pm(k/N;N) = (N^+/N+1+N^-/N)\,\frac{(1-\vartheta_\pm/N)(N^\pm + k)}{N^\pm + N-k + (1-\vartheta_\pm/N)\,(N^\mp+k)}.$$
Therewith we obtain the expansion of the rates w.r.t. $1/N$, 
$$ 
 f^\pm(x;N) = x + \frac 1 N g^\pm(x)+{\cal O}(N^{-2}),\qquad  g^\pm(x) = \vartheta_\pm x^2 - \vartheta_\pm x + N^\pm.
$$
Consequently, we obtain 
%
\begin{eqnarray*}
G^\pm(x) 
&=& \int^x\frac {g^\pm(y)}{y}\, dy 
= \frac 1 2 \vartheta_\pm x^2 - \vartheta_\pm x + N^\pm\,\ln(x)
= -\frac 1 2 \vartheta_\pm \bigg(2x-x^2\bigg) + N^\pm\,\ln(x)
\end{eqnarray*}
and the stationary distribution (we use eqn.~\eqref{weakHamiltonian} with $\kappa_\pm=0$) 
\begin{eqnarray}
\varphi(x) &=& C \,\,x^{N^+-1}\,(1-x)^{N^--1}\,\,
e^{-\vartheta_+ x(2-x)/2 }\,\,
e^{-\vartheta_-(1-x)(2-(1-x))/2}.
\end{eqnarray}
If we compare the stationary distribution of the weak effects q-voter model and the weak effects reinforcement model, we find a striking similarity: In both cases, the measure is an adaptation of the beta distribution, 
$ \varphi(x) = C \,\,x^{N^+-1}\,(1-x)^{N^--1}\,\,
e^{a_+\zeta(x)}\,\,
e^{a_-\zeta(1-x)}$ where $a_\pm=\kappa_\pm$ and $\zeta(x)=-x(1-\ln(x))$ in the q-voter case, while $a_\pm=\vartheta_\pm/2$ and $\zeta(x)=-x(2-x)$ 
in the reinforcement case. As both functions $\zeta(x)$ resemble each other in that $\zeta(0)=0$, $\zeta(1)=1$, and both are convex, we expect very similar behavior for the two models if we take $\vartheta_\pm=2\kappa_\pm$ (also inspect Figure~\ref{bifuFig}).\\
As a last remark, we note that we do allow in the weak q-voter model and the weak reinforcement model not only for positive parameter values $\kappa_\pm$ and $\vartheta_\pm$ but also for negative values. While positive values for these parameters lead to filter bubbles and echo chambers (a person hesitates to change her mind), negative values have the interpretation that the person are open-minded and pay particular attention to the opposite opinion. In the case of the functioning of democracies and its institutions, open-minded people are much more preferable because it is easier to find compromises for solving political problems.

\subsection{Model behavior}
We will not go deeper into the analysis (which can be found, e.g. for the Curie-Weiss model in~\cite{Nicolao2019a,Krapivsky2017}, for the q-voter model in the strong effects limit in~\cite{Castellano2009qVoter} and for the reinforcement model in~\cite{Mueller2022}), but simply refer to Figure~\ref{bifuFig}, which shows that all four models undergo a phase transition if the coupling between the individuals is sufficiently large. We emphasize that also the models based on the voter- and zealot model (which {\it per se} do not allow for phase transitions), the two kinds of q-voter model and the reinforcement model, exhibit phase transitions. The mechanisms modifying the effects of zealots target on in/outgroup communication. If in/outgroup communication is sufficiently strong, a bimodal distribution appears via a phase transition. Also the behavior under non-symmetric conditions (parameters) leads to similar behavior of all four models.\\
It is interesting to note that the Curie-Weiss and the strong effects q-voter model incorporate the population size $N$ explicitly, while the (weak effects limit of) the q-voter and the reinforcement model become independent of $N$. Usually, in applications, $N$ is very large, and if we naively take $N$ to the population size, the variance generated by the model is much smaller than the variance that is present in empirical data. The way out is to assume that individuals cluster together and to estimate an effective population size $N=N_{eff}$ along with the other parameters, which of course is slightly dubious, but pragmatic~\cite{Nicolao2019a}. The weak effects models elegantly circumvent this difficulty.

\section{Data analysis}

We use data from four different Western democracies which represent different types of government and electoral systems (see~\cite[pp 145--161, 271]{Hague2016} and~\cite{Reynolds2005} 
and Table~\ref{tabDesign} for further details). Concerning the governmental system we have one presidential (US), one semi-presidential (France), and two parliamentary systems (UK, Germany). In the case of the electoral systems we have two majority systems with a first-past-the-post design and relative majority (US, UK) and one majority system in France with a two-round system and absolute majority. In Germany we have a mixed member proportional system that combines both a first-past-the-post vote and proportional representation. Finally, for each country we study different numbers of elections (US: six presidential elections, 2000-2020; UK: 20 parliamentary elections, 1945-2019; FRA: second round of five presidential elections, 2002-2022; GER: two parliamentary elections, 2017-2021; the data sources are indicated in the data availability statement). 
\begin{table}
\begin{center}{\small 
\begin{tabular}{p{2cm}|p{2cm}|p{2cm}|p{2cm}|p{2cm}}
 & US & UK & France & Germany\\
 \hline 
 Government system & presidential	& parliamentary	& semi-presidential & parliamentary\\
 \hline
 Electoral system & 
 Plurality/ majority: single-member districts, first-past-the-post, relative majority	
 &Plurality/ majority: single-member districts, first-past-the-post, relative majority	
 &Plurality/ majority: two-round system, absolute majority	 &Mixed: mixed member proportional\\
 \hline 
 Election years covered in the analysis	&2000-2020 (6~presidential elections)	& 1945-2019 (20~parliamentary elections)	&2002-2022 (5~presidential elections, second round)	&2017-2021 (2~parliamentary elections)
\end{tabular}}
\end{center}
\caption{Design of the study: data set used.}\label{tabDesign}
\end{table}
Such a design is useful to understand how the models used can explain the dynamics in different institutional settings with different political cultures and in varying periods of time. This comparative approach gives us more information about the functioning of the mechanisms in different contexts (e.g.~\cite{Fortunato2007,Chatterjee2013})  and contributes to the existing research often based on single-case studies (e.g.~\cite{Galam2017, Palombi2014}).  
In the analysis, we consider each election district as an i.i.d.\ repetition of the election. Herein we obviously reduce the complexity of the data by neglecting social co-factors and spatial effects. In that, we obtain an empirical distribution of vote shares and can use a maximum likelihood estimator. Please find the technical details, particularly the algorithm used to perform the maximum-likelihood estimation, in Appendix~\ref{dataAna}.\\ 
The present approach to data analysis is based on a steady-state assumption, that is, the opinion formation process is assumed to be approximately in equilibrium. If there is a huge shift in the vote share of a candidate or party in recent times, this assumption steady-state might not be met. The tables with estimates, p-values for the Kolmogorov-Smirnov test, and AICs  can be found in Appendix~\ref{dataAna}. \par\medskip

\begin{figure}
\begin{center}
\includegraphics[width=10cm]{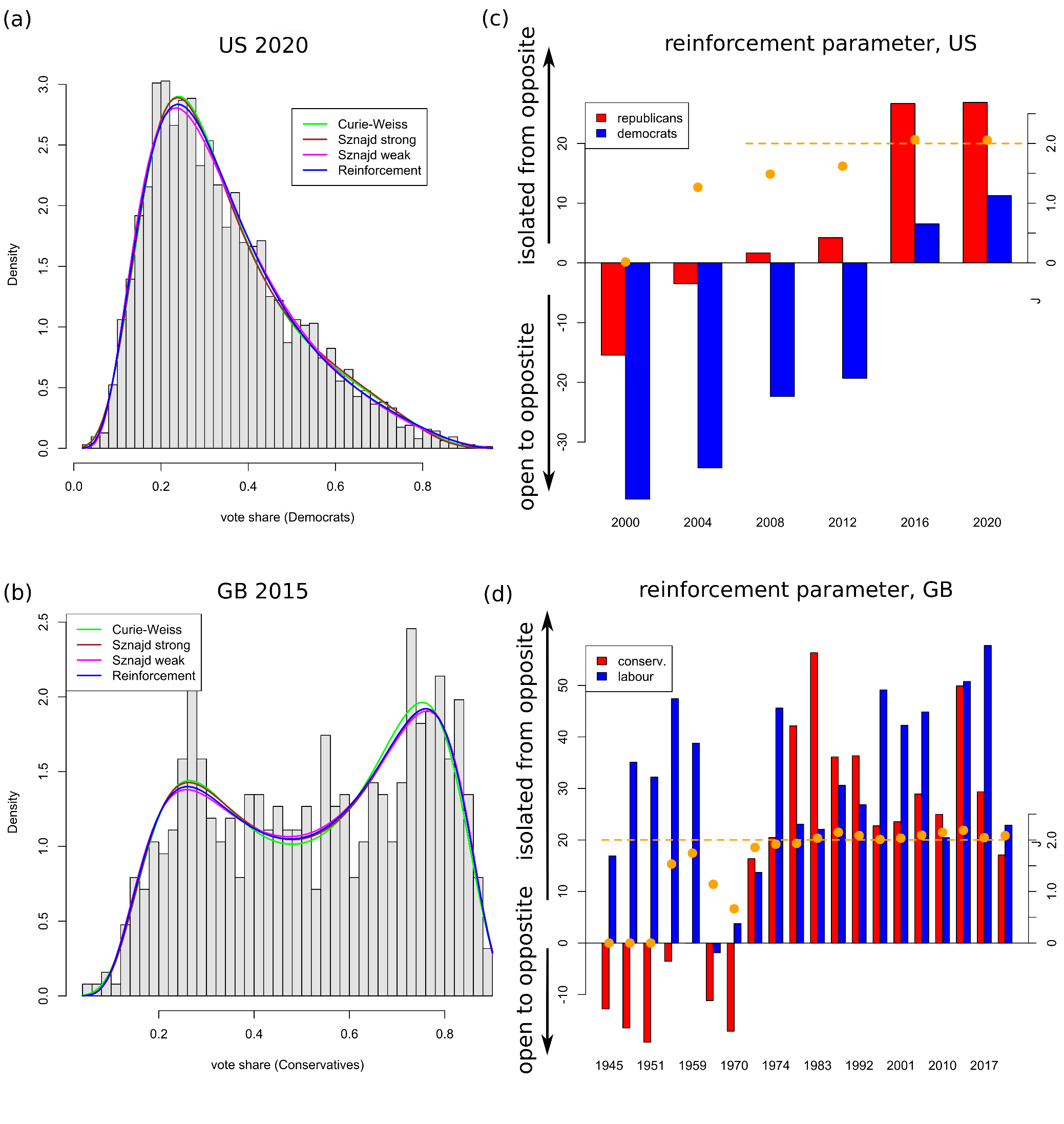}
\end{center}
\caption{(a), (b): Histograms of the vote shares for (a) the presidential US elections in 2020 (b) the parliamentary elections in 2015 in UK, together with the probability density for the four models. (c), (d):  Bars indicate the reinforcement parameters $\vartheta_\pm$ for (c) democrats and republicans and (d) the conservatives and the Labour Party (left axis); orange bullets indicate parameter $J$ of the Curie-Weiss model (right axis), while the horizontal dashed orange line indicates the threshold for the phase transition of that model for $h=0$.}\label{us:gb:data}
\end{figure}

United States data: The densities of the four models (Curie-Weiss, weak and strong q-voter model and reinforcement model) are rather similar (Fig.~\ref{us:gb:data} (a)), and also the Kolmogorov-Smirnov tests resemble each other (see table in Section~\ref{usAppendix}). Only in the year 2000, the p-values of this test are small (between $0.018$ and $0.04$); in that year, apart from the democrats and the republicans, also the green candidate did win a small but reasonable fraction of votes, such that the dichotomous models might not be completely suited. In 2016, we have also about four percent of third-party votes, but at that point, the models fit better. Maybe this is due to the fact, that polarization in the American society has already grown during these past 16 years. The AIC selects always the (weak) reinforcement model as best-suited model but in the year 2020, where the weak q-voter model fits best. However, in this year the reinforcement model and the weak q-voter model are very close. The strong models always perform worse 
(Fig.~\ref{US:AIC}). We clearly find a trend in the parameters which shows that the model moves in time more and more towards a phase transition. 
\par\medskip

\par\medskip 
United Kingdom data: The trend in the reinforcement parameters/coupling is particularly interesting  (Fig.~\ref{us:gb:data}, (d)). As can be clearly visualized in the empirical and the estimated distributions for the election from 2015, the UK indeed became super-critical. We observe a bimodal distribution in 2015. It is most interesting to see that the theoretical prediction of possible phase transitions  is realized in the UK.\par\medskip

\begin{figure}
\begin{center}
\includegraphics[width=\textwidth]{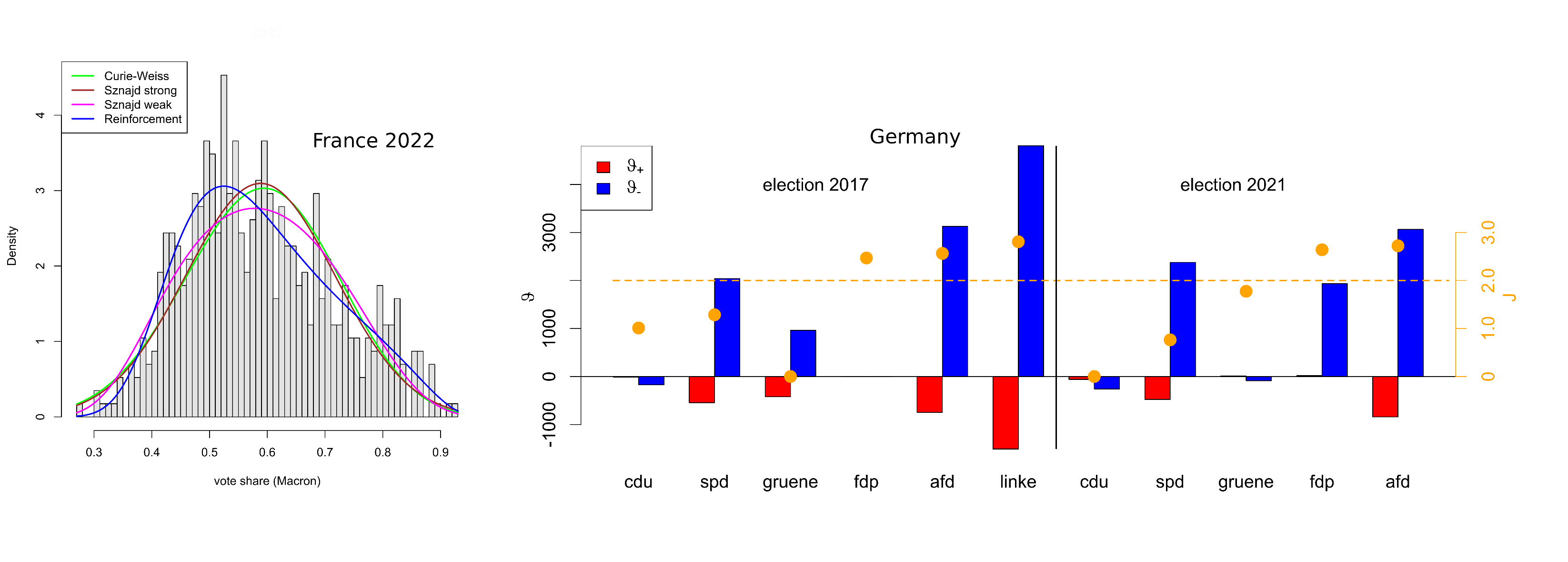}
\end{center}
\caption{Right: Histogram of the vote shares for the presidential elections in France (2022), together with the probability density for the four models. Left: Histogram of $\vartheta_\pm$, and $J$ (together with critical level, orange) for the parties in 2017, 2021 which hold more than 5\% of the votes (threshold), apart of the csu, which is a Bavarian local.}\label{Fr:Ger:Data}
\end{figure}

France data: Particularly in 2022, the empirical distribution of vote shares is screwed. The strong effects models have difficulties dealing with this result, while the weak effects models are more flexible; particularly the reinforcement model still performs well. 
\par\medskip

German data: As we have a proportional electoral system, the dichotomous 
model requires adaptation: For each party, we distinguish between the votes in favor of this party versus the votes for all other parties. In Fig.~\ref{Fr:Ger:Data} we find the reinforcement parameters together with the coupling $J$ of the Curie-Weiss model for those parties in the 2017 and 2021 elections that did receive at least 5\% of the votes (where we did disregard the CSU, as this is a Bavarian local party which  only stands for election in few election districts). If we focus on the best models (lowest AIC), these models indeed are able to meet the empirical vote share distributions  quite well (Kolmogorov-Smirnov test), with only one exception: The left-wing party Left Party (die linke) in the election 2017. Due to historical reasons, this party performs very differently in the federal states coming from the former East respectively West Germany. Though also these historical effects can be understood to be based on opinion dynamics and in/out-group behavior, all models have difficulties to capturing this data structure. The assumption of a homogeneously mixed population may no longer be appropriate; instead, a two-island model would capture the communication structure better. Also the (relatively recent) right-wing party AfD, which also performs rather differently in the two regions (former East/West Germany), poses a problem for all models but the reinforcement model, at least according to the Kolmogorov-Smirnov test. However, the difference in the two regions is less pronounced than in the case of the Left Party, which might also be the reason why the reinforcement model is still able to handle the vote share data of the AfD.
\par\medskip

\begin{figure}
\begin{center}
\includegraphics[width=8cm]{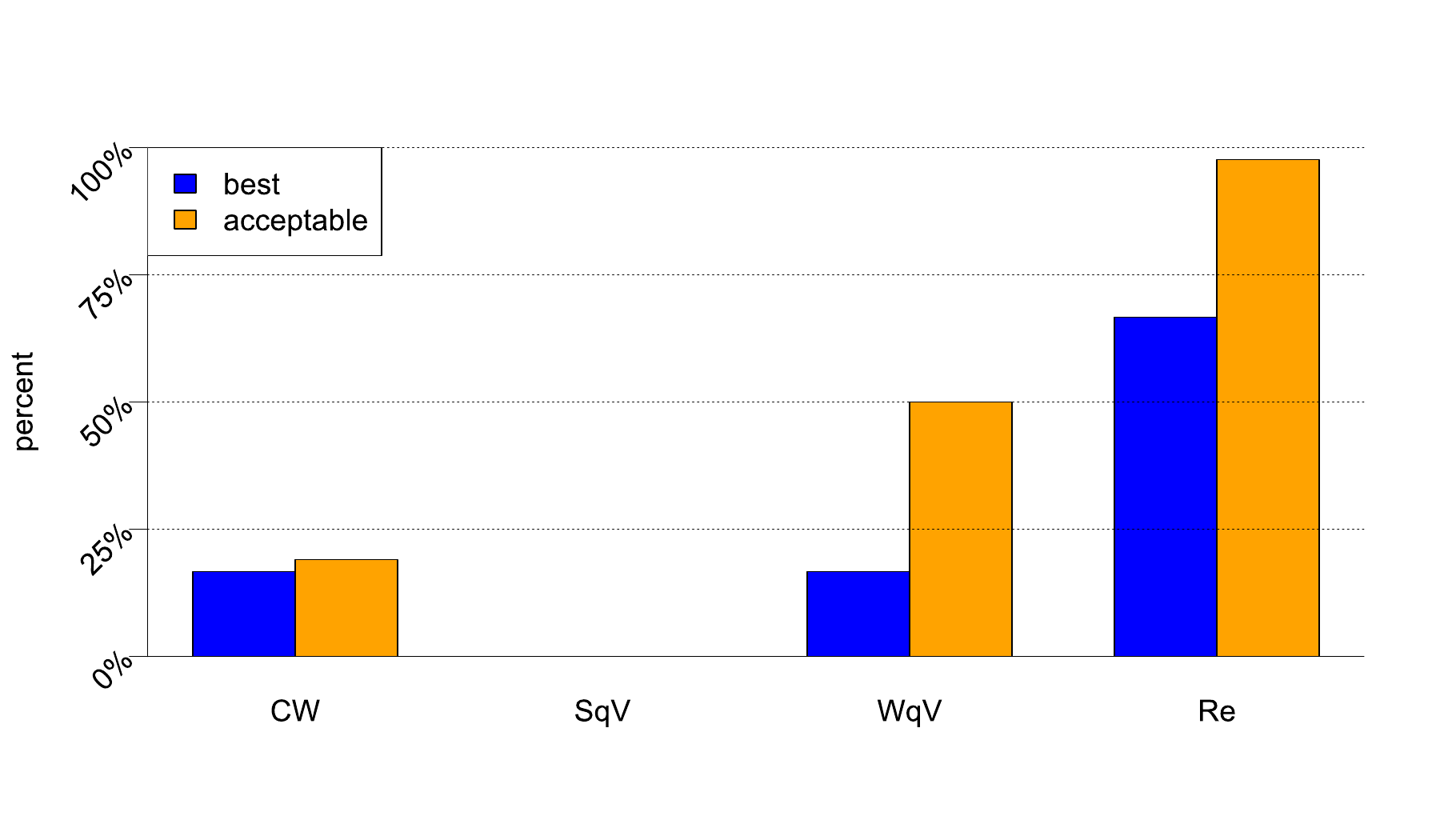}%
\parbox{2cm}{\vspace*{-5cm}
\parbox{5cm}{\scriptsize 
\begin{tabular}{ll}CW& Curie-Weiss\\[2pt]
SqV& strong effects q-voter\\[2pt]
WqV& weak effects q-voter\\[2pt]
Re& reinforcement
\end{tabular}
}}
\end{center}
\caption{Performance of the four models. For the 42 elections we consider in the present paper, we indicate the percentage for which the models performed best according to the AIC (yellow), and acceptable (AIC has a maximum distance of 2 to the best model).}\label{perform}
\end{figure}

Summary of the estimations: In almost all elections, at least some of the models describe the empirical data adequately (according to the Kolmogorov-Smirnov test). If we compare the performance of the models according to the AIC, we find that the weak effects models outperform the strong effects models, and the weak effects reinforcement model is superior to the weak effects q-voter model (Fig.~\ref{perform}). It is interesting, that -- though the structure of the weak effects models are very similar -- nevertheless we find a difference in their suitability for practical applications.\\ 
We should keep in mind that we work with aggregated data (only the outcome of election in election districts which typically have a population of hundreds of thousands of individuals). We might consider the model parameters as a reduction of the data complexity to a low-dimensional parameter space, which allows us to better interpret the data. As the Kolmogorov-Smirnov test indicates the appropriateness of the models, we can be confident that the models capture at least some fundamental structure in the data. In that, the interpretation of the parameters suggested by the models will be appropriate. 

\subsection{Political Science Interpretations}
\par\medskip 
United States: Looking at the parameters measuring the actual strength of reinforcement in  the reinforcement model, we can see that since the year 2000, there has been a continuous trend for a much more polarized voting behavior in the United States. For the voters of the Republican party, the switch could be seen already in 2008 with the election of Barack Obama and it continued in his re-election while it became dominant in 2016 and 2020 when Donald Trump became candidate of the Republican party. The voters for the Democratic party clearly also changed from open- to closed-mindedness in the political realm. In fact, what we can see is that during 20 years the political discourse became so polarized in the US public arena that now both the parties and their voters are becoming clearly more and more separated from each other. As Binder~\cite{Binder2015}  for example shows is that the frequency of legislative gridlock has risen since the 1990s and other studies show that polarization in the US citizenry is not going down~\cite{PewResearchCenter2024}. 
\par\medskip 

Great Britain: Due to the electoral system (first-past-the-post) and the governmental system (parliamentary) it is clearly useful for the two dominant parties – the Labour Party on the left and the Conservative Party (Tories) on the right – to keep a certain or even a high degree of polarization so that they can form the government on their own. The bars of the reinforcement model show this clearly for Labour for almost all elections since 1945, but for the Tories, this strategy started only in the 1970s and became very dominant since the second election in 1974 and the Thatcher years 1979-1990. Additionally, we can see that in 2019 the reinforcement parameters had not been so strong. This might be the result of the Brexit decision in 2016 and its political aftermath in which the Conservatives had become the party of the Leavers (in support of Brexit) and the Remainers have split between Labour and the Liberal Democrats~\cite{Prosser2020}. 


\par\medskip

France: The French Party System has undergone a political change in the last years since 2017 so some already argue that we may see the rise of a new French Party System~\cite{Gougou2017}. In 2017, both presidential candidates of the traditional left-wing (Parti Socialiste) and right-wing parties (Les Républicains) were disqualified in the first round of the presidential elections~\cite{Durovic2019}. And this happened again in 2022, when both Emmanuel Macron of the centrist "La R\'epublique En Marche" and Marine Le Pen of the new-named right-wing populist ``Rassemblement National'' were for the second time the political opponents in the second round of the election~\cite{Durovic2022}. That said we can see in Figure~\ref{France:AIC} for the previous elections the models didn't explain so much for the second round.



\par\medskip 
Germany: Here we can see that the right-wing populist party Alternative for Germany (AfD) was able in both elections to create their own space for resonating with their voters. In 2017, both the Social Democrats (SPD), the Left Party, and the Greens had been able to mobilize their voters against the AfD, but in 2021 it was the SPD and the liberal party (FDP). Compared to the left parties, the FDP tried to position itself as the party of Freedom where some parts of the party raised also their critical voice against the political means the former government of the two conservative parties CDU and CSU and the Social Democrats used during the Covid-19 pandemic. This can be observed in their parliamentary work where they used so called ``Kleine Anfragen'' to question the government parties the most compared to all the other remaining opposition parties in the German Bundestag~\cite{Donovan2021}. Therefore, they had been a competitor both to the AfD and the other left and center parties and gained also a lot of support by young voters~\cite{Faas2022}.



\section{Discussion of the findings and their interpretation in political sciences}
In the present paper, we first discussed the connection between Potts models and stochastic opinion models, both for well mixing populations. Particularly, we did provide an alternative approach to the q-voter model as a  natural extension of the Zealot model in the view of the Glauber dynamics. Consequently, motivated by similar constructions in population genetics, we introduced a strong and particularly also a weak effects continuum limit. While the strong continuum limit is generically possible (and approximates locally a normal distribution), the weak effects limit requires additional structure. In that, the Curie-Weiss model only allows for  the strong limit, while the q-voter model has both limits. Afterwards, we additionally introduced the reinforcement model, which has its foundation not in the Potts machinery but is derived based on considerations about the impact of the several kinds of cognitive biases on communication, especially on the resulting in/out-group communication strategies. Also that model allows for both continuum limits, where it turned out that the weak q-voter and the weak reinforcement model are very similar in their mathematical structure. Basically both are an adaptation of the beta distribution, which is the well-known weak effects limit of the zealot (or Moran) model~\cite{Ewens2004}. We also found that only models that are based on the voter model allow for a weak effect limit, which indicates that these models are in some sense special.\\
After these theoretical considerations, we turned to test the models based on election data. Herein, we used each election district as an i.i.d.\ repetition of the election, neglecting social co-factors which vary between election districts, as well as spatial factors. We also assume the opinion process to be approximately in equilibrium, such that the stationary distribution is an adequate description of the data; in case of a large shift in the vote share of a candidate or party, also this assumption can be called into question. Though the simplicity of the approach, mostly we found that the models meet the data quite well, where the weak models performed better than the strong models, and the weak reinforcement model outperformed the weak q-voter model. It is interesting that the weak effect models seem to be better suited to describe the data appropriately. The comparison of models and data always is challenging, but it seems that also in population genetics, where the weak effects as weak selection are often used, rather supports that these kind of models are well suited for real-world applications~\cite{Loewe2010}. The background could be that striking and immediately disruptive events are rare. Most stimuli are weak and require time to unfold their effect. If this observation is correct, weak effect models with their slow time scale might indeed be a better description of reality than strong effect models with a fast time scale. As a practical consequence, we propose to use in empirical studies rather weak effects opinion models than strong effects models, which also has the advantage that we do not need to choose an appropriate population size, which is a well-known problem in itself~\cite{Ewens2004,Nicolao2019a}. Though the models allowing for a weak effect limit are rather special, they seem to be a powerful description of reality and still have sufficient flexibility to address different mechanisms and different real-world (electorate) systems. 
\par\medskip

Elections are at the centre stage of modern representative democracies. Correspondingly, research on elections and attempts to explain the formation of their results are also central. Prominent and established approaches use statistical data concerning the social characteristics of voters to determine their voting behavior (e.g.~\cite{Lipset1967}). But there is an ongoing debate among scholars, that the correlation between social characteristics and voting behavior has diminished over the last decades (e.g.~\cite{Dalton1984,Karvonen2001,Kitschelt2015}). Additionally, party membership is also in decline which has also consequences for voter turnout and voting behavior (e.g.~\cite{Biezen2014,Siaroff2009}). As Clarke et al.~\cite{Clarke2004}  bluntly declared, for understanding electoral choice one has to look elsewhere. It is not that the classic approaches have lost all their explanatory power, but it makes sense to look for explanations that are less context dependent. \\
By focusing on the dynamics of opinion formation preceding the act of voting, the models discussed in this paper promise insights both into the empirical explanation of elections as such as well as important aspects of the theory of democracy.\\
Our leading assumption, ensuring a larger independence of specific social contexts, is opinion formation via frequently contacting social sources of information constituting a ubiquitous mechanism of collective decision-making. For sure, this assumption also holds for elections. The sources of information here may be real persons or media of all kinds. Pamphlets, newspapers, magazines, radio, TV, and the diversity of social media have accompanied political discussion since the early modern period. Albeit the basic mechanism is taken to be the same everywhere, its effects may be modulated by the impact of or the interaction with other mechanisms of other sections within an “organized complexity “. Electoral systems in their specific forms are nested within the broader construction of a political system. The effects of opinion formation processes concerning elections are shaped by specific institutional settings as well as the political culture of the respective countries. \\
The model’s reduction to only two opinions may look like to crude a simplification. But it is not such implausible as it may appear at first glance. As Denver and Johns~\cite{Denver2020}  stress, when preparing their decisions, voters don’t “sit down before an election to comb through the parties’ manifestos and make detailed calculations of the costs and benefits of voting for each party. (…) Such a process is neither realistic nor particularly rational~\cite[p.294]{Denver2020}”. Rather voters base their decision on just a few subjects dominating the discussion. “Issue Voting” and “Valence Voting” denominate the approaches based on that assumption. Where Issue Voting stresses the main issues of the election campaign like economic questions or social policy, Valence Voting focuses for example on the performance of the incumbent government. “Neither involves complex calculations; indeed, the simpler versions of both approaches have fared better when confronted with the empirical evidence (ibid.)”. And both aspects fare better than explaining results with respect to voters’ social characteristics. Therefore, looking at opinion and opinion formation in this simple form offers a promising starting point to delve deeper into campaigning and voting.\\
In this view, the models in their present form may be interpreted as to suppose a stage of the election process, where the main issues are already settled. From here the model may be enhanced by stepwise nesting the basic opinion formation process within a whole set of similar modeled ones. The determination of the salient issues and valences may be the next step. The party that succeeds in putting its topics in place may be an advantage. A variant of the Issue Voting approach stresses, that it is not alone the preferences on the specific topics of the very election of today that affect the voter’s decision but general values and principles~\cite[p.295]{Denver2020}. We can think of processes concerning ideological backgrounds running on a larger time scale spanning across two or more elections affecting the probability with which a voter makes up his mind. Another perspective would look at the developments within the zealots in particular. This would mean looking at the development of party programmes and strategies also in the form of opinion formation among party members. The possibilities are manifold.\\
The reinforcement model in particular also highlights important aspects concerning the theory of democracies. Especially in the liberal tradition of democracy, it is a common view to interpret campaigning and elections as a market analogous competition, where votes are exchanged for programmes and personal (see~\cite{Dahl1972}). Competition appears as a form of regulated and so limited conflict. The opponent’s purpose is not to harm the antagonist, but only to be better. The idea behind this is, that the aspiration to trump the adversary leads to the advancement of a common good, qualitatively better, or cheaper products and processes in economy, better theories and methods in science, and better programmes and personnel in politics.\\
The zealot model, as a predecessor of the reinforcement model, was used before in economics for market analysis~\cite{Kirman1993}. But behind this application lies a model designed to explain foraging processes of ant colonies~\cite{Pasteels_1987}. It describes a form of collective information processing against an uncertain environment. Time and again the colony has to leave an established feeding ground and look for another in time. It is inspiring to see the similarities. Political communities also have to alter their processes and organization because of altering circumstances, for example transforming their way of living to a more sustainable way. In this way an open political process, defining problems and looking for solutions is a collective information processing, too. This idea was emphasized by Karl Popper~\cite{Popper2020}, for example, and further developed by John Dryzek~\cite{Dryzek1990}. It is not implausible to assume, that a part of the success of democracies in general is their dealing with the world’s shakiness in an analogous way to modern science.\\

Whatever makes the workers of the ant colony change their paths, the driving force behind the parliamentary process, at least from the perspective of the liberal standard model of democracy, is party competition. Since party competition is itself affected by special interests and personal ambition of politicians, democracies need additional features to balance these forces and bring the wanted effects of the competition to the fore. This has been part of considerations from Harrington~\cite{Harrington1992} to Tocqueville~\cite{Tocqueville2000} to Dewey~\cite{Dewey2016} but shall not be the point here.\\
What the reinforcement model enables us to see is the possible polarization between the opponents. It is important to note, that polarization indicating a higher grade of conflict is not a problem per se as well as higher grades of conflict aren’t. As sociologist Lewis Coser (1967)~\cite{Coser1967} argued for, conflicts in the first place point to societal problems within a society urging to deal with their causes. If the community is productively addressing that challenge, society reintegrates on a new level. We could see such effects for example in the course of the environmental movement in the 70s to 90s in Germany. The polarization on the side of the Greens was high in the beginning, when they cracked open the consensus of the established parties on the use of nuclear power, and became lower again, when environmental issues were successfully established on the agenda.\\
Polarization may become problematic when the reinforcement effects are strong on both sides of the debate. Conflicts can be disruptive too. American philosopher of law Ronald Dworkin was asking already in 2006~\cite{Dworkin2006} looking at the polarization in the US “Is Democracy possible here?”. The polarization in the US is not in decline since then (Pew Research Center 2022, \cite{PewResearchCenter2024}). Polarization may also become problematic when actors show no tendency to consent or compromise enabling reintegration. This appears to be the case with the populist movements and parties of the last decade. On the other hand, party polarization may generate stronger party attachments, which could also be a desirable strategy for political competitors~\cite[p. 350]{Lupu2014}. And probably here is the point, where political scientists (at least at the moment) have to reach out for other methods than mathematical modeling, too. Qualitative analysis of texts or focus groups may be an appropriate means here.
However, it should be ascertained that the reinforcement model supplies us with a strong indicator concerning an important variable of political processes. And since the claim that a society is polarized is also often used in an alarmist way, impeding compromise, the more it is helpful to have this indicator. And it should be ascertained further that because of its context independence, the model will be useful, when we look not alone at well-established democracies of the West but also on young ones or democracies in other world regions.

\vspace{4cm}

\noindent
{\bf Data availability}\par\medskip 
\noindent 
\underline{United States Data.}\\
\url{https://dataverse.harvard.edu/dataset.xhtml?persistentId=doi:10.7910/DVN/VOQCHQ       }\\
All data from the US have been accessed at 6-24-2021.\par\medskip 
\noindent 
 \underline{France data.}\\
 2002-2012:\\
 \url{https://www.data.gouv.fr/fr/posts/les-donnees-des-elections/}\\[4pt]
2017:\\
\url{https://www.data.gouv.fr/fr/datasets/election-presidentielle-des-23-avril-et-7-mai-2017-resultats-definitifs-du-2nd-tour/}\\[4pt]
2022:\\
\url{https://www.data.gouv.fr/fr/datasets/election-presidentielle-des-10-et-24-avril-2022-resultats-definitifs-du-2nd-tour/}\\
All France data are accessed at 12-8-2023.\par\medskip 

\noindent 
\underline{United Kingdom data.}\\
\url{https://commonslibrary.parliament.uk/research-briefings/cbp-8647}\\
All UK data are accessed at 11-5-2021.\par\medskip 

\noindent 
\underline{German data.}\\
\url{https://www.bundeswahlleiter.de/en/bundeswahlleiter.html}\\
The  German data for the 2017 election have been  accessed at 7-8-2018, that for the 2021 election at 10-5-2022.\par\medskip 

\newpage

\bibliographystyle{abbrv}
\bibliography{socialSciencesLit,politcs}

\begin{appendix}
\setcounter{section}{0}
\section{Data analysis}\label{dataAna}
The maximum likelihood estimates have been performed in R~\cite{R}, based on the optimizer {\tt nlm}. In case of the Curie-Weiss and the strong q-voter model, all parameters have been optimized in one run.\\ 
The weak effects models have two parameters related to the zealots ($N^\pm$ respectively $\eta^\pm$), and two parameters related to ingroup/outgroup communication structures ($\lambda_\pm$, $\kappa_\pm$, and $\vartheta_\pm$). Here, we used an EM-like algorithm by alternately optimizing the parameters for the zealots and holding the in/outgroup parameters, and then optimizing the parameters for in/outgroup while holding the zealot parameters, until the vector of parameters becomes approximately constant.\par\medskip 

The Kolmogorov-Smirnov test is performed to test for the appropriateness of the current model, and the AIC is determined in order to compare the models. The columns of the tables are to read specifically for the model, e.g. the $J/\eta^+/N^+$ column holds $J$ in case of the Curie-Weiss model, $\eta^+$ for the strong q-voter model, and $N^+$ for the weak q-voter model and reinforcement model.\par\medskip 

In the resulting tables, we mark the models that have the minimal AIC (best performing models) for the corresponding elections in yellow. Of course, if the AIC of two models is close, it is not sensible to dismiss one of the two models. Burnham et al.~\cite{Burnham2004} suggest that models which have a difference of at most two do perform comparably. These models are marked in gay. 
Additionally, we provide a graphic for the differences in the AIC's. Here, we use for each given election the AIC of the Curie-Weiss model as a reference, and provide 
$$ \Delta \mbox{AIC} = \mbox{AIC other model}-\mbox{AIC Curie-Weiss}.$$
That is, if this value is positive, the ``other model'' performs better than the Curie-Weiss model, if it is negative, the Curie-Weiss model performs better. Moreover, if $\Delta \mbox{AIC}$ is e.g.\ larger for the weak q-voter model than for the reinforcement model, then the weak q-voter model performs better than the reinforcement model.

\newpage 
\subsection{United States data}\label{usAppendix}
\begin{figure}[ht!]
\begin{center}
    \includegraphics[width=6cm]{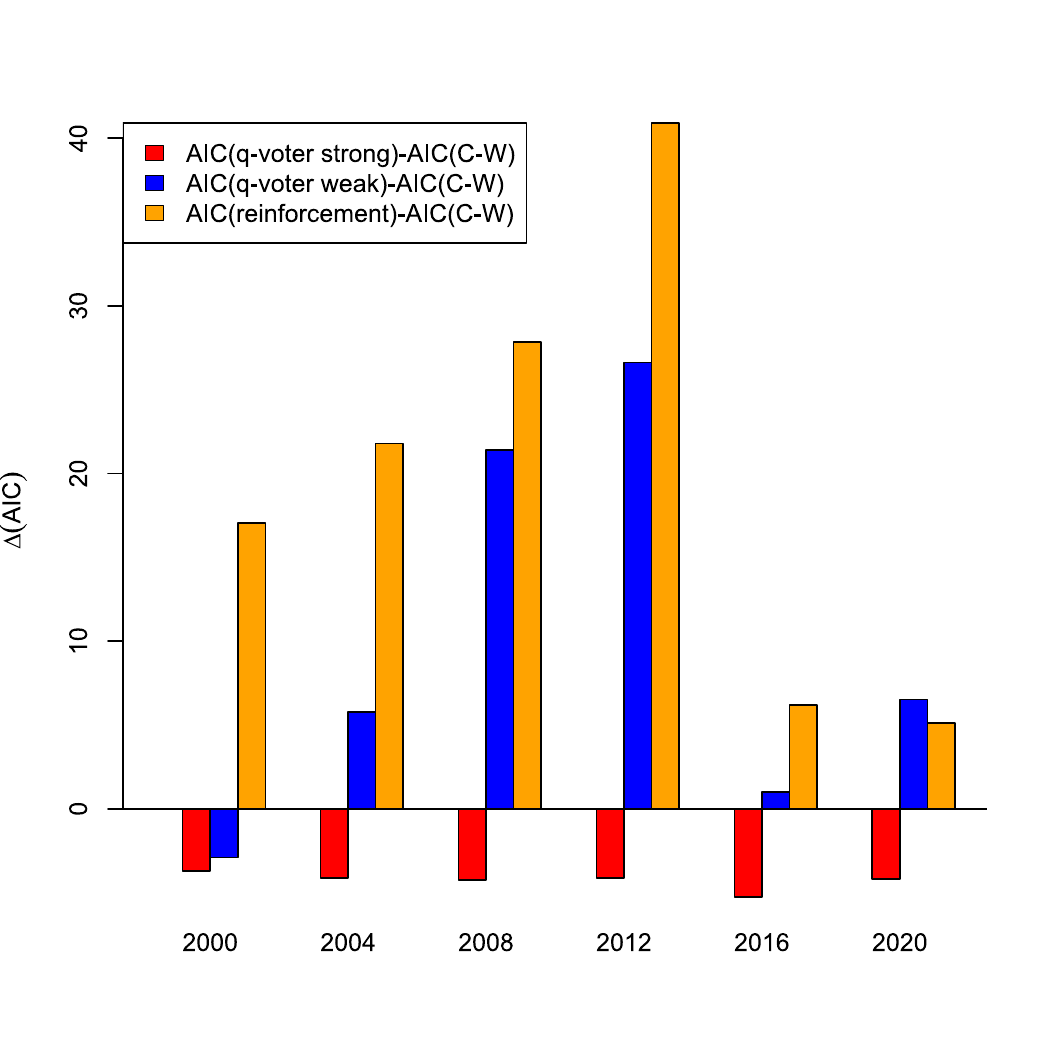}
\end{center}
\caption{Comparison of the models' AICs for the US.}\label{US:AIC}
\end{figure}

US presidential elections in the years 2000-2020 are analyzed, where we only consider democrats and republicans and the vote share of the democrats (e.g. votes for democrats/(votes for democrats or republicans), and dismiss the votes for any other candidate. The performance of the AICs for the models is shown in Fig.~\ref{US:AIC}.
\par\medskip 
{\small 

\adjustbox{max width=\textwidth}{
\begin{tabular}{ll|lllll|ll}
 year & model & J/$\eta^+$/$N^+$ & h/$\eta^-$/$N^-$ & $\lambda_+$/$\kappa_+$/$\vartheta_+$ & $\lambda_-$/$\kappa_-$/$\vartheta_-$ & N & AIC & $p_{KS}$\\
 \hline
 2000 &Curie-Weiss& 0.015 & -0.38 & $-$ & $-$ & 15.1 & -4299.2 & 0.04 \\
 2000 &q-voter strong& 7 & 19.5 & -2.7 & -12.7 & 12.2 & -4295.5 & 0.046 \\
 2000 &q-voter weak& 3.7 & 0.00031 & -3.2 & -22.1 & $-$ & -4296.3 & 0.023 \\
 \rowcolor{yellow}
 2000 &reinforcement& 3 & 6.9e-06 & -15.5 & -39.6 & $-$ & -4316.3 & 0.1 \\
 \hline
 2004 &Curie-Weiss& 1.3 & -0.2 & $-$ & $-$ & 31.8 & -4150.6 & 0.39 \\
 2004 &q-voter strong& 31.9 & 50.2 & 36.6 & 69.4 & 31.5 & -4146.4 & 0.43 \\
 2004 &q-voter weak& 6.8 & 2.8e-06 & 7.5 & -21.2 & $-$ & -4156.4 & 0.29 \\
 \rowcolor{yellow}
 2004 &reinforcement& 4.4 & 0.12 & -3.5 & -34.3 & $-$ & -4172.4 & 0.48 \\
 \hline
 2008 &Curie-Weiss& 1.5 & -0.11 & $-$ & $-$ & 34.2 & -3489.9 & 0.31 \\
 2008 &q-voter strong& 15.8 & 20.3 & 23.3 & 32.7 & 35 & -3485.6 & 0.23 \\
 2008 &q-voter weak& 7.9 & 0.04 & 11.1 & -17.3 & $-$ & -3511.3 & 0.49 \\
 \rowcolor{yellow}
 2008 &reinforcement& 5 & 1.3 & 1.7 & -22.4 & $-$ & -3517.7 & 0.28 \\
 \hline
 2012 &Curie-Weiss& 1.6 & -0.13 & $-$ & $-$ & 31.7 & -3074.7 & 0.15 \\
 2012 &q-voter strong& 22 & 25.2 & 35.5 & 42.8 & 31.8 & -3070.6 & 0.19 \\
 2012 &q-voter weak& 6.5 & 0.018 & 10.5 & -14.7 & $-$ & -3101.3 & 0.3 \\
 \rowcolor{yellow}
 2012 &reinforcement& 4.3 & 0.92 & 4.2 & -19.3 & $-$ & -3115.6 & 0.88 \\
 \hline
 2016 &Curie-Weiss& 2.1 & -0.085 & $-$ & $-$ & 56.8 & -3167.7 & 0.71 \\
 2016 &q-voter strong& 4.9 & 8 & 10.1 & 19.3 & 58.2 & -3162.5 & 0.79 \\
 2016 &q-voter weak& 9.3 & 2.4 & 22.3 & -5 & $-$ & -3168.7 & 0.3 \\
 \rowcolor{yellow}
 2016 &reinforcement& 6.4 & 4.1 & 26.7 & 6.5 & $-$ & -3173.9 & 0.59 \\
 \hline
 2020 &Curie-Weiss& 2.1 & -0.084 & $-$ & $-$ & 55.7 & -3066 & 0.51 \\
 2020 &q-voter strong& 13.8 & 25.3 & 25.7 & 59.7 & 56.6 & -3061.8 & 0.59 \\
 \rowcolor{yellow}
 2020 &q-voter weak& 9.2 & 3.5 & 21.5 & -2 & $-$ & -3072.5 & 0.84 \\
 \rowcolor{lightgray}
 2020 &reinforcement& 6.5 & 5.2 & 26.9 & 11.3 & $-$ & -3071.1 & 0.69 \\
 \hline
 \end{tabular}
}
}
\par\medskip 

\eject

\newpage 
\subsection{UK data}\label{ukAppendix}
\begin{figure}[h!]
\begin{center}
    \includegraphics[width=10cm]{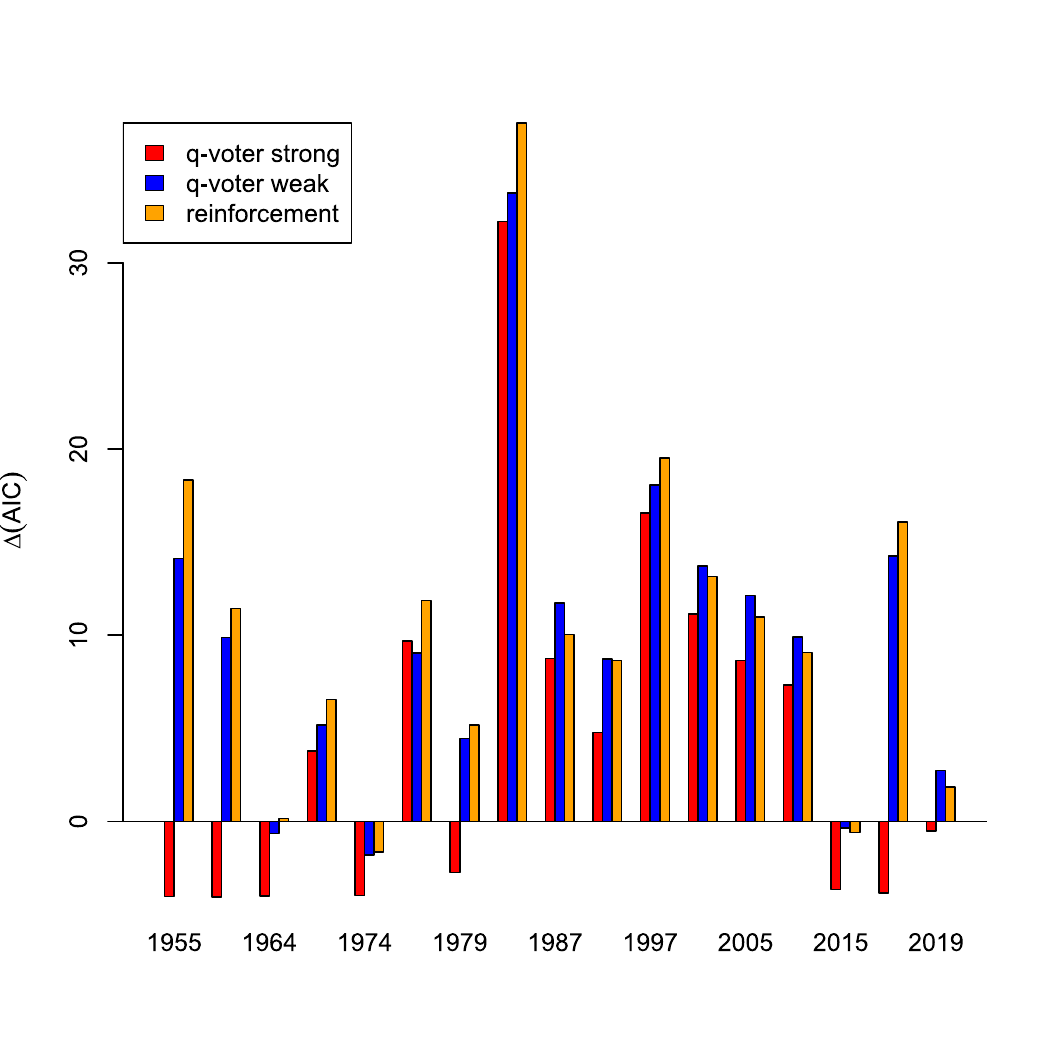}
\end{center}
\caption{Comparison of the models' AICs for the UK.}\label{GB:AIC}
\end{figure}

We investigate the vote share of the conservative among the conservative and Labour votes.  The performance of the AICs for the models is shown in Fig.~\ref{GB:AIC}.
\par\medskip 
{\small 

\adjustbox{max width=\textwidth}{
 \begin{tabular}{ll|lllll|ll}
 year & model & J/$\eta^+$/$N^+$ & h/$\eta^-$/$N^-$ & $\lambda_+$/$\kappa_+$/$\vartheta_+$ & $\lambda_-$/$\kappa_-$/$\vartheta_-$ & N & AIC & $p_{KS}$\\
 \hline
 1945 &Curie-Weiss& 3.3e-08 & -0.26 & $-$ & $-$ & 12.1 & -593.4 & 0.86 \\
 1945 &q-voter strong& 7.6 & 0.67 & -6.1 & -0.56 & 9.5 & -589.6 & 0.84 \\
 1945 &q-voter weak& 1.1 & 11.2 & -13.7 & 10.6 & $-$ & -592.6 & 0.16 \\
 \rowcolor{yellow}
 1945 &reinforcement& 3.3 & 11.7 & -12.8 & 16.9 & $-$ & -596.7 & 0.64 \\
 \hline
 1950 &Curie-Weiss& 7.1e-08 & -0.039 & $-$ & $-$ & 12.2 & -628.5 & 0.64 \\
 1950 &q-voter strong& 4.2 & 4.1 & -24 & -23 & 3.7 & -624.9 & 0.61 \\
 1950 &q-voter weak& 0.098 & 21.3 & -21.7 & 40.3 & $-$ & -643.4 & 0.98 \\
 \rowcolor{yellow}
 1950 &reinforcement& 2.8 & 15 & -16.5 & 35.1 & $-$ & -646.2 & 0.96 \\
 \hline
 1951 &Curie-Weiss& 6.4e-07 & 0.013 & $-$ & $-$ & 13.1 & -652.9 & 0.34 \\
 1951 &q-voter strong& 2.6 & -0.0066 & -9.4 & -0.68 & 6.5 & -651.4 & 0.45 \\
 1951 &q-voter weak& 3.3e-08 & 29.3 & -26.1 & 60.6 & $-$ & -666.4 & 0.92 \\
 \rowcolor{yellow}
 1951 &reinforcement& 2.6 & 14.4 & -19.3 & 32.2 & $-$ & -669.4 & 0.94 \\
 \hline
 1955 &Curie-Weiss& 1.5 & 0.018 & $-$ & $-$ & 43.8 & -715.8 & 0.65 \\
 1955 &q-voter strong& 17.9 & 16.5 & 28.5 & 25.5 & 43.4 & -711.8 & 0.74 \\
 1955 &q-voter weak& 1.7 & 17.7 & -16.5 & 31.9 & $-$ & -730 & 0.96 \\
 \rowcolor{yellow}
 1955 &reinforcement& 5.3 & 16.8 & -3.5 & 47.4 & $-$ & -734.2 & 0.87 \\
 \hline
 1959 &Curie-Weiss& 1.7 & 0.022 & $-$ & $-$ & 63.2 & -699.5 & 0.47 \\
 1959 &q-voter strong& 27.2 & 24 & 49.4 & 41.4 & 61.6 & -695.4 & 0.52 \\
 \rowcolor{lightgray}
 1959 &q-voter weak& 3.3 & 17.9 & -11.8 & 33.4 & $-$ & -709.4 & 0.82 \\
 \rowcolor{yellow}
 1959 &reinforcement& 5.7 & 14 & 0.018 & 38.8 & $-$ & -710.9 & 0.88 \\
 \hline
 \rowcolor{lightgray}
 1964 &Curie-Weiss& 1.1 & -0.011 & $-$ & $-$ & 22.4 & -639 & 0.72 \\
 1964 &q-voter strong& 9.6 & 9.5 & 11.3 & 11.1 & 21.9 & -635 & 0.75 \\
 \rowcolor{lightgray}
 1964 &q-voter weak& 2 & 5.5 & -9.1 & 0.89 & $-$ & -638.4 & 0.87 \\
 \rowcolor{yellow}
 1964 &reinforcement& 3.1 & 5.3 & -11.2 & -1.9 & $-$ & -639.2 & 0.95 \\
 \hline
 1970 &Curie-Weiss& 0.66 & 0.017 & $-$ & $-$ & 15.1 & -625.1 & 0.41 \\
 1970 &q-voter strong& -0.62 & 0.56 & 0.73 & 2.2 & 67.9 & -628.9 & 0.96 \\
 \rowcolor{lightgray}
 1970 &q-voter weak& 0.35 & 10 & -15.3 & 13.8 & $-$ & -630.3 & 0.91 \\
 \rowcolor{yellow}
 1970 &reinforcement& 1.9 & 6.5 & -17.1 & 3.8 & $-$ & -631.7 & 0.89 \\
 \hline
 \rowcolor{yellow}
 1974 &Curie-Weiss& 1.9 & 0.0039 & $-$ & $-$ & 50.5 & -497.5 & 0.7 \\
 1974 &q-voter strong& 5.7 & 5.6 & 11.6 & 11.2 & 50.6 & -493.5 & 0.69 \\
 \rowcolor{lightgray}
 1974 &q-voter weak& 8.2 & 7.7 & 10 & 8.8 & $-$ & -495.7 & 0.65 \\
 \rowcolor{lightgray}
 1974 &reinforcement& 7.3 & 6.6 & 16.4 & 13.7 & $-$ & -495.9 & 0.74 \\
 \hline
 1974 &Curie-Weiss& 1.9 & -0.018 & $-$ & $-$ & 51.7 & -450.8 & 0.15 \\
 1974 &q-voter strong& 0.042 & 0.22 & 1 & 1.5 & 178.1 & -460.5 & 0.88 \\
 1974 &q-voter weak& 7.1 & 14.2 & 6.8 & 26.8 & $-$ & -459.8 & 0.34 \\
 \rowcolor{yellow}
 1974 &reinforcement& 7.1 & 13.2 & 20.5 & 45.6 & $-$ & -462.6 & 0.89 \\
 \hline
 1979 &Curie-Weiss& 1.9 & 0.021 & $-$ & $-$ & 57.5 & -473.5 & 0.31 \\
 1979 &q-voter strong& 10.4 & 3.8 & 25.4 & 6.7 & 59.7 & -470.8 & 0.37 \\
 \rowcolor{lightgray}
 1979 &q-voter weak& 18.4 & 7.8 & 37 & 7.1 & $-$ & -478 & 0.63 \\
 \rowcolor{yellow}
 1979 &reinforcement& 12.6 & 7.7 & 42.2 & 23.1 & $-$ & -478.7 & 0.65 \\
 \hline
 1983 &Curie-Weiss& 2 & 0.072 & $-$ & $-$ & 32.8 & -327.7 & 0.044 \\
 1983 &q-voter strong& 1.4 & 0.04 & 5 & 0.81 & 108.1 & -360 & 0.54 \\
 1983 &q-voter weak& 22.2 & 5.3 & 51.4 & 3.8 & $-$ & -361.5 & 0.51 \\
 \rowcolor{yellow}
 1983 &reinforcement& 14.2 & 5.6 & 56.3 & 22 & $-$ & -365.3 & 0.67 \\
\end{tabular}
}

\adjustbox{max width=\textwidth}{
\begin{tabular}{ll|lllll|ll}
year & model & J/$\eta^+$/$N^+$ & h/$\eta^-$/$N^-$ & $\lambda_+$/$\kappa_+$/$\vartheta_+$ & $\lambda_-$/$\kappa_-$/$\vartheta_-$ & N & AIC & $p_{KS}$\\
 \hline
 1987 &Curie-Weiss& 2.1 & 0.029 & $-$ & $-$ & 51.9 & -293.6 & 0.21 \\
 1987 &q-voter strong& 0.016 & 0.011 & 1 & 1 & 692.9 & -302.4 & 0.61 \\
 \rowcolor{yellow}
 1987 &q-voter weak& 11.1 & 8.6 & 23.6 & 17.9 & $-$ & -305.4 & 0.55 \\
 \rowcolor{lightgray}
 1987 &reinforcement& 8.6 & 6.7 & 36.1 & 30.6 & $-$ & -303.7 & 0.57 \\
 \hline
 1992 &Curie-Weiss& 2.1 & 0.017 & $-$ & $-$ & 52.5 & -309.1 & 0.2 \\
 1992 &q-voter strong& 0.2 & 0.099 & 1.5 & 1.2 & 114.1 & -313.9 & 0.44 \\
 \rowcolor{yellow}
 1992 &q-voter weak& 12.3 & 8.1 & 25.8 & 14.3 & $-$ & -317.8 & 0.47 \\
 \rowcolor{lightgray}
 1992 &reinforcement& 9.2 & 6.7 & 36.3 & 26.8 & $-$ & -317.7 & 0.6 \\
 \hline
 1997 &Curie-Weiss& 2 & -0.056 & $-$ & $-$ & 40.4 & -376 & 0.34 \\
 1997 &q-voter strong& 0.07 & 1.8 & 0.83 & 6 & 96.3 & -392.6 & 0.52 \\
 \rowcolor{lightgray}
 1997 &q-voter weak& 6.2 & 16 & 7.3 & 33.8 & $-$ & -394.1 & 0.9 \\
 \rowcolor{yellow}
 1997 &reinforcement& 6.2 & 13.1 & 22.8 & 49.1 & $-$ & -395.5 & 0.54 \\
 \hline
 2001 &Curie-Weiss& 2 & -0.044 & $-$ & $-$ & 44.6 & -363.2 & 0.48 \\
 2001 &q-voter strong& 0.039 & 0.15 & 1 & 1.4 & 183.1 & -374.3 & 0.22 \\
 \rowcolor{yellow}
 2001 &q-voter weak& 6.7 & 15.2 & 9.1 & 32.2 & $-$ & -376.9 & 0.37 \\
 \rowcolor{lightgray}
 2001 &reinforcement& 6.2 & 11.3 & 23.5 & 42.3 & $-$ & -376.3 & 0.22 \\
 \hline
 2005 &Curie-Weiss& 2.1 & -0.018 & $-$ & $-$ & 56.5 & -316.5 & 0.2 \\
 2005 &q-voter strong& 0.052 & 0.13 & 1.1 & 1.3 & 173.7 & -325.2 & 0.17 \\
 \rowcolor{yellow}
 2005 &q-voter weak& 7.9 & 16.6 & 12.2 & 37.1 & $-$ & -328.7 & 0.21 \\
 \rowcolor{lightgray}
 2005 &reinforcement& 7.1 & 11 & 28.9 & 44.9 & $-$ & -327.5 & 0.18 \\
 \hline
 2010 &Curie-Weiss& 2.1 & 0.021 & $-$ & $-$ & 39.4 & -211.1 & 0.061 \\
 2010 &q-voter strong& 0.0093 & 0.0073 & 1 & 1 & 709.6 & -218.4 & 0.15 \\
 \rowcolor{yellow}
 2010 &q-voter weak& 7.7 & 6.1 & 15.6 & 11.9 & $-$ & -221 & 0.12 \\
 \rowcolor{lightgray}
 2010 &reinforcement& 6.2 & 4.9 & 25 & 20.4 & $-$ & -220.1 & 0.17 \\
 \hline
 \rowcolor{yellow}
 2015 &Curie-Weiss& 2.2 & 0.0076 & $-$ & $-$ & 83.4 & -307.9 & 0.71 \\
 2015 &q-voter strong& 1.7 & 1.6 & 4.9 & 4.4 & 86.9 & -304.2 & 0.71 \\
 \rowcolor{lightgray}
 2015 &q-voter weak& 13.8 & 14.1 & 30.8 & 32.4 & $-$ & -307.5 & 0.79 \\
 \rowcolor{lightgray}
 2015 &reinforcement& 10.5 & 10.4 & 49.9 & 50.8 & $-$ & -307.3 & 0.79 \\
 \hline
 2017 &Curie-Weiss& 2 & 0.0039 & $-$ & $-$ & 76 & -416.3 & 0.47 \\
 2017 &q-voter strong& 4.9 & 5.1 & 10.8 & 11.6 & 77 & -412.5 & 0.5 \\
 \rowcolor{lightgray}
 2017 &q-voter weak& 8.7 & 17.8 & 10.9 & 38.8 & $-$ & -430.6 & 0.45 \\
 \rowcolor{yellow}
 2017 &reinforcement& 8.2 & 14.4 & 29.4 & 57.7 & $-$ & -432.4 & 0.61 \\
 \hline
 2019 &Curie-Weiss& 2.1 & 0.036 & $-$ & $-$ & 45.5 & -314.8 & 0.77 \\
 2019 &q-voter strong& 0.043 & 0.065 & 1.1 & 1.2 & 143.9 & -314.2 & 0.57 \\
 \rowcolor{yellow}
 2019 &q-voter weak& 6.3 & 7.6 & 9.2 & 15.5 & $-$ & -317.5 & 0.8 \\
 \rowcolor{lightgray}
 2019 &reinforcement& 5.3 & 5.9 & 17.1 & 22.9 & $-$ & -316.6 & 0.77 \\
 \hline
 \end{tabular}
}

}
\par\medskip 

\eject
\newpage
$\qquad$

\newpage
\subsection{France data}\label{franceAppendix}
\begin{figure}[h!]
\begin{center}
    \includegraphics[width=8cm]{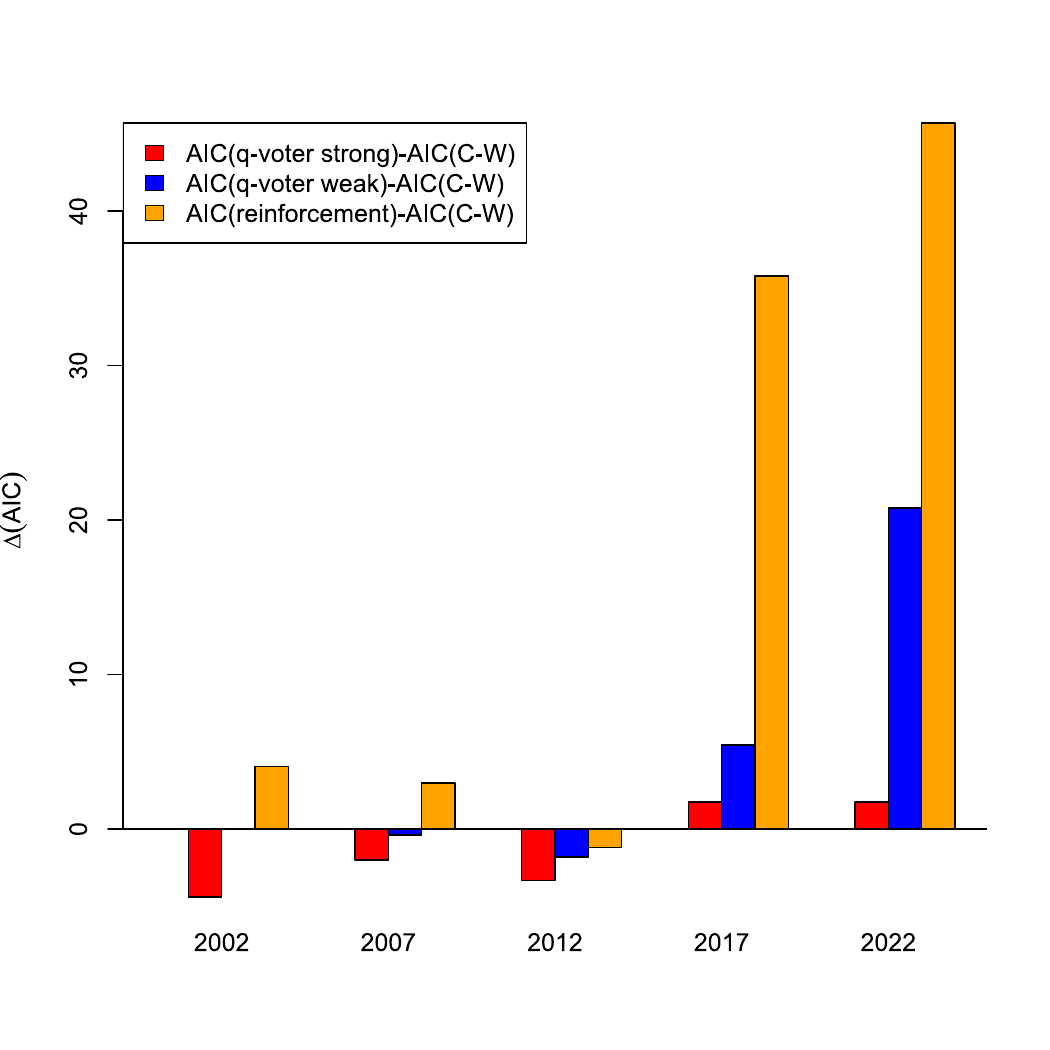}
\end{center}
\caption{Comparison of the models' AICs for France.}\label{France:AIC}
\end{figure}

This is the second round of the presidential election, and we consider the vote share of the winning candidate. The performance of the AICs for the models is shown in Fig.~\ref{France:AIC}.
\par\medskip 

\adjustbox{max width=\textwidth}{
\begin{tabular}{ll|lllll|ll}
 year & model & J/$\eta^+$/$N^+$ & h/$\eta^-$/$N^-$ & $\lambda_+$/$\kappa_+$/$\vartheta_+$ & $\lambda_-$/$\kappa_-$/$\vartheta_-$ & N & AIC & $p_{KS}$\\
 \hline
 2002 &Curie-Weiss& 0.95 & 0.98 & $-$ & $-$ & 58 & -1672.5 & 0.11 \\
 2002 &q-voter strong& 35.5 & 6 & 14 & 2 & 47 & -1668.1 & 0.099 \\
 2002 &q-voter weak& 23.5 & 5 & -56.5 & -7 & $-$ & -1672.5 & 0.16 \\
 \rowcolor{yellow}
 2002 &reinforcement& 533 & 9 & 1275.5 & -452.7 & $-$ & -1676.5 & 0.32 \\
 \hline
 2007 &Curie-Weiss& 0.097 & 0.11 & $-$ & $-$ & 41.8 & -1297.7 & 0.58 \\
 2007 &q-voter strong& 3.2 & 2.9 & -280.7 & -241.6 & 1.1 & -1295.7 & 0.66 \\
 2007 &q-voter weak& 41.1 & 5.3e-05 & 56.2 & -57 & $-$ & -1297.3 & 0.48 \\
 \rowcolor{yellow}
 2007 &reinforcement& 23.2 & 6.1e-05 & -0.32 & -86.2 & $-$ & -1300.7 & 0.29 \\
 \hline
 \rowcolor{yellow}
 2012 &Curie-Weiss& 1.4e-07 & 0.082 & $-$ & $-$ & 29.9 & -1138 & 0.2 \\
 2012 &q-voter strong& 2.8 & 2.6 & -162.4 & -145.5 & 1.2 & -1134.7 & 0.23 \\
 \rowcolor{lightgray}
 2012 &q-voter weak& 0.006 & 0.00031 & -33.9 & -30.4 & $-$ & -1136.2 & 0.22 \\
 \rowcolor{lightgray}
 2012 &reinforcement& 0.54 & 0.0014 & -66.8 & -64.1 & $-$ & -1136.8 & 0.24 \\
 \hline
 2017 &Curie-Weiss& 1.2e-07 & 0.74 & $-$ & $-$ & 14.3 & -819.2 & 0.18 \\
 2017 &q-voter strong& 8.1 & 4.5 & -817 & -343.8 & 0.41 & -821 & 0.25 \\
 2017 &q-voter weak& 0.064 & 1 & -27.6 & -8.8 & $-$ & -824.7 & 0.24 \\
 \rowcolor{yellow}
 2017 &reinforcement& 109.6 & 6.3 & 325.2 & -61.2 & $-$ & -855 & 0.95 \\
 \hline
 2022 &Curie-Weiss& 1.5e-09 & 0.38 & $-$ & $-$ & 13.4 & -718.8 & 0.0077 \\
 2022 &q-voter strong& 0.009 & 0.029 & -206.6 & -129.5 & 0.085 & -720.5 & 0.003 \\
 2022 &q-voter weak& 85.1 & 0.11 & 192 & -49.2 & $-$ & -739.6 & 0.097 \\
 \rowcolor{yellow}
 2022 &reinforcement& 65.6 & 7.4 & 211.9 & -17.2 & $-$ & -764.5 & 0.96 \\
 \hline
 \end{tabular}
}

\par\medskip 

\eject

\newpage 
\subsection{German data}\label{gerAppendix}
\begin{figure}[h!]
\begin{center}
    \includegraphics[width=12cm]{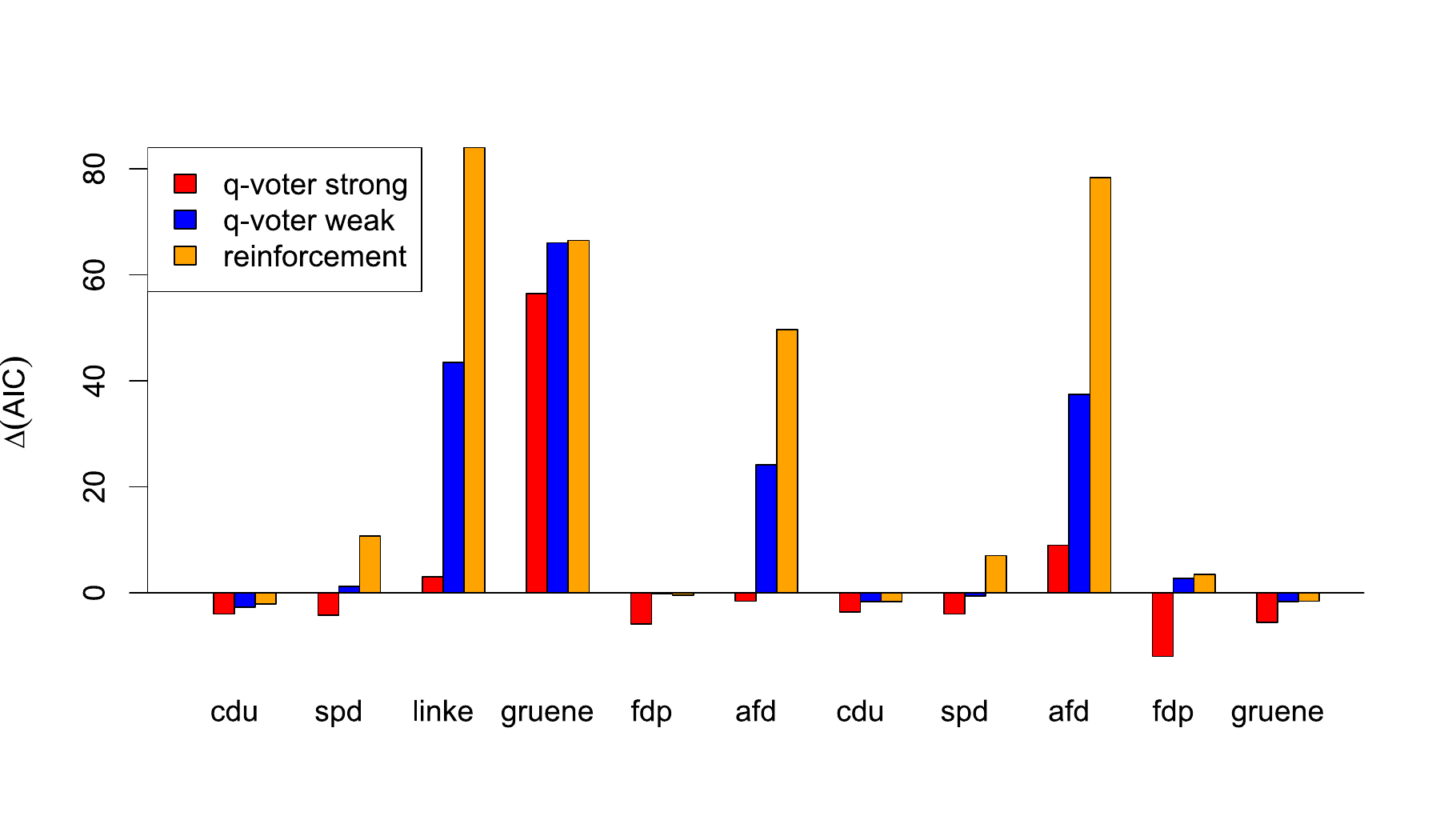}
\end{center}
\caption{Comparison of the models' AICs for Germany.}\label{GER:AIC}
\end{figure}

We include all parties which did reach a vote share of at least 5\%, but the CSU, which is a local party and thus only stands for election in few districts. Note that "die linke" is present in the parliament of 2021 though this party did not reach 5\%, and in that, we did exclude this party in 2021. In order to fit our dichotomy model, we focus on the vote share of the focal party, essentially distinguishing between supporters of this party, and supporters of any  other party. The performance of the AICs for the models is shown in Fig.~\ref{GER:AIC}.
\par\medskip 
{\small 

\adjustbox{max width=\textwidth}{
\begin{tabular}{lll|lllll|ll}
 year & model & Party& J/$\eta^+$/$N^+$ & h/$\eta^-$/$N^-$ & $\lambda_+$/$\kappa_+$/$\vartheta_+$ & $\lambda_-$/$\kappa_-$/$\vartheta_-$ & N & AIC & $p_{KS}$\\
 \hline
 \rowcolor{yellow}
 2017 &Curie-Weiss& cdu & 1 & -0.41 & $-$ & $-$ & 123.1 & -745.2 & 0.81 \\
 2017 &q-voter strong& cdu & 22.6 & 48.9 & 19.8 & 57.9 & 123.4 & -741.2 & 0.81 \\
 2017 &q-voter weak& cdu & 14 & 26.7 & -15 & -56.3 & $-$ & -742.5 & 0.87 \\
 2017 &reinforcement& cdu & 14.4 & 9.4e-05 & -14.7 & -173 & $-$ & -743.1 & 0.83 \\
 \hline
 2017 &Curie-Weiss& spd & 1.3 & -0.63 & $-$ & $-$ & 68.7 & -814.1 & 0.23 \\
 2017 &q-voter strong& spd & 7.2 & 31.3 & 5 & 33.1 & 55.5 & -809.8 & 0.17 \\
 2017 &q-voter weak& spd & 10.1 & 0.27 & 18.9 & -80.6 & $-$ & -815.3 & 0.18 \\
 \rowcolor{yellow}
 2017 &reinforcement& spd & 19.6 & 752.9 & -545.5 & 2039.8 & $-$ & -824.8 & 0.4 \\
 \hline
 2017 &Curie-Weiss& linke & 2.8 & -0.11 & $-$ & $-$ & 81.9 & -1068.6 & 3.6e-08 \\
 2017 &q-voter strong& linke & 0.06 & 20.5 & 0.6 & 82 & 108.6 & -1071.6 & 2.7e-07 \\
 2017 &q-voter weak& linke & 9.1 & 2.7 & 42.8 & 9.3 & $-$ & -1112.1 & 5.7e-07 \\
 \rowcolor{yellow}
 2017 &reinforcement& linke & 14.2 & 1800.1 & -1509.8 & 4806 & $-$ & -1152.6 & 6e-05 \\
 \hline
 2017 &Curie-Weiss& gruene & 2e-06 & -2.8 & $-$ & $-$ & 19.2 & -1047.3 & 2.6e-06 \\
 2017 &q-voter strong& gruene & 0.95 & 45.2 & 0.72 & 12.5 & 50.8 & -1103.8 & 0.26 \\
 \rowcolor{lightgray}
 2017 &q-voter weak& gruene & 4.7 & 0.063 & 15.3 & -164.1 & $-$ & -1113.4 & 0.7 \\
 \rowcolor{yellow}
 2017 &reinforcement& gruene & 4.4 & 472.5 & -421.4 & 961.3 & $-$ & -1113.8 & 0.71 \\
 \hline
 \rowcolor{yellow}
 2017 &Curie-Weiss& fdp & 2.5 & -0.21 & $-$ & $-$ & 244.3 & -1320.7 & 0.5 \\
 2017 &q-voter strong& fdp & 11.9 & 116.1 & 1.3 & 8.4 & 131.5 & -1314.9 & 0.34 \\
 \rowcolor{lightgray}
 2017 &q-voter weak& fdp & 17.2 & 5.4 & 53.7 & -308.8 & $-$ & -1320.5 & 0.57 \\
 \rowcolor{lightgray}
 2017 &reinforcement& fdp & 14 & 117.1 & 0.14 & 1.1 & $-$ & -1320.2 & 0.57 \\
 \hline
 2017 &Curie-Weiss& afd & 2.6 & -0.088 & $-$ & $-$ & 110 & -985.3 & 0.0082 \\
 2017 &q-voter strong& afd & 1.1 & 23 & 2.2 & 86.7 & 117.5 & -983.7 & 0.011 \\
 2017 &q-voter weak& afd & 14.5 & 8.9 & 54.5 & 23.7 & $-$ & -1009.4 & 0.094 \\
 \rowcolor{yellow}
 2017 &reinforcement& afd & 22.4 & 1081.9 & -747.5 & 3127.1 & $-$ & -1035 & 0.77 \\
 \hline
 \rowcolor{yellow}
 2021 &Curie-Weiss& cdu & 2.6e-07 & -1.3 & $-$ & $-$ & 64 & -778.3 & 0.6 \\
 2021 &q-voter strong& cdu & 16.6 & 53 & -40 & -215.2 & 30.8 & -774.6 & 0.75 \\
 \rowcolor{lightgray}
 2021 &q-voter weak& cdu & 1.8 & 8.8 & -31 & -156.9 & $-$ & -776.6 & 0.76 \\
 \rowcolor{lightgray}
 2021 &reinforcement& cdu & 3.5 & 0.8 & -60.7 & -261.1 & $-$ & -776.6 & 0.75 \\
 \hline
 2021 &Curie-Weiss& spd & 0.76 & -0.7 & $-$ & $-$ & 72.8 & -829.6 & 0.74 \\
 2021 &q-voter strong& spd & 13.1 & 38.5 & 5.3 & 21.2 & 63 & -825.5 & 0.74 \\
 2021 &q-voter weak& spd & 13.1 & 1.2 & 14 & -97.2 & $-$ & -828.9 & 0.74 \\
 \rowcolor{yellow}
 2021 &reinforcement& spd & 38.7 & 827.5 & -477 & 2376.9 & $-$ & -836.6 & 0.93 \\
 \hline
 2021 &Curie-Weiss& afd & 2.7 & -0.12 & $-$ & $-$ & 71.9 & -971.4 & 1.3e-05 \\
 2021 &q-voter strong& afd & 0.044 & 14.9 & 0.59 & 56.4 & 122.8 & -980.4 & 7.7e-05 \\
 2021 &q-voter weak& afd & 8.6 & 4.1 & 37.3 & 11 & $-$ & -1008.8 & 0.0014 \\
 \rowcolor{yellow}
 2021 &reinforcement& afd & 13.7 & 1092.2 & -841 & 3064.4 & $-$ & -1049.7 & 0.21 \\
 \hline
 2021 &Curie-Weiss& fdp & 2.6 & -0.032 & $-$ & $-$ & 430.3 & -1424.6 & 0.37 \\
 2021 &q-voter strong& fdp & 19.7 & 170.3 & 1.5 & 11.5 & 198.3 & -1412.6 & 0.15 \\
 \rowcolor{lightgray}
 2021 &q-voter weak& fdp & 40.9 & 254 & 74.7 & 703.9 & $-$ & -1427.3 & 0.64 \\
 \rowcolor{yellow}
 2021 &reinforcement& fdp & 43.3 & 522.6 & 19 & 1933.9 & $-$ & -1428 & 0.79 \\
 \hline
 \rowcolor{yellow}
 2021 &Curie-Weiss& gruene & 1.8 & -0.61 & $-$ & $-$ & 52.2 & -813.8 & 0.18 \\
 2021 &q-voter strong& gruene & 1.2 & 24.7 & 0.98 & 21.2 & 36.2 & -808.2 & 0.081 \\
 \rowcolor{lightgray}
 2021 &q-voter weak& gruene & 3.8 & 22.1 & -2.1 & -21.3 & $-$ & -812.1 & 0.13 \\
 \rowcolor{lightgray}
 2021 &reinforcement& gruene & 3.8 & 0.3 & 13.9 & -87.8 & $-$ & -812.2 & 0.13 \\
 \hline
 \end{tabular}
}

}
\eject

\end{appendix}

\end{document}